\documentclass[3p, gulliver, 12pt, A4]{elsarticle}

\usepackage[utf8]{inputenc}
\usepackage{amsmath}
\usepackage[english]{babel}
\usepackage{mathtools}
\usepackage{amsfonts}

\usepackage{appendix}

\usepackage{amssymb}
\usepackage{mathrsfs}
\usepackage{textcomp}
\usepackage{gensymb}

\usepackage{graphicx}
\usepackage[scriptsize]{caption}
\usepackage{caption}
\usepackage{subcaption}

\usepackage{wrapfig}

\usepackage{listings}

\usepackage{tikz}
\usepackage{tikz-3dplot}
\usetikzlibrary{arrows}
\usepackage{xparse}
\usetikzlibrary{3d,shapes,shadings}

\usepackage{enumitem}

\usepackage{footmisc}

\usepackage{soul}

\usepackage{transparent}

\usepackage{eso-pic}

\usepackage{pdfpages}
\usepackage{float}

\usepackage{hyperref}
\hypersetup{
	colorlinks=true,
	citecolor=Purple,
	urlcolor=MyPink,      
	linkcolor= Blue,
	menucolor = Green,
}

\usepackage[makeroom]{cancel}
 \usepackage{wasysym}
 
\usepackage[font=footnotesize,labelfont=bf]{caption}

\begin{document}
\begin{frontmatter}

	\title{Reinforcement Learning for Adaptive Time-Stepping in the Chaotic Gravitational Three-Body Problem}

	\author[1]{Veronica Saz Ulibarrena
	\corref{cor1}\fnref{fn1}}
	\cortext[cor1]{Corresponding author}
	\ead{ulibarrena@strw.leidenuniv.nl
	}

	\author[1]{Simon~Portegies~Zwart}
	\ead{spz@strw.leidenuniv.nl}
	
	\address[1]{Leiden Observatory, Leiden University, Einsteinweg 55, 2333 CC, Leiden, The Netherlands}

	\begin{abstract}
	Many problems in astrophysics cover multiple orders of magnitude in spatial and temporal scales. While simulating systems that experience rapid changes in these conditions, it is essential to adapt the (time-) step size to capture the behavior of the system during those rapid changes and use a less accurate time step at other, less demanding, moments. We encounter three problems with traditional methods. Firstly, making such changes requires expert knowledge of the astrophysics as well as of the details of the numerical implementation. Secondly, some parameters that determine the time-step size are fixed throughout the simulation, which means that they do not adapt to the rapidly changing conditions of the problem. Lastly, we would like the choice of time-step size to balance accuracy and computation effort. 
	We address these challenges with Reinforcement Learning by training it to select the time-step size dynamically. We use the integration of a system of three equal-mass bodies that move due to their mutual gravity as an example of its application. With our method, the selected integration parameter adapts to the specific requirements of the problem, both in terms of computation time and accuracy while eliminating the expert knowledge needed to set up these simulations.
	Our method produces results competitive to existing methods and improve the results found with the most commonly-used values of time-step parameter. This method can be applied to other integrators without further retraining. We show that this extrapolation works for variable time-step integrators but does not perform to the desired accuracy for fixed time-step integrators.  
	\end{abstract}
	
	
\end{frontmatter}

\section{Introduction}
\label{Introduction}

Reinforcement Learning (RL) has recently been growing in popularity for multiple applications. Based on the positive results of RL as a control mechanism, we explore its use for the simulation of the motion of celestial bodies due to their mutual gravity. In these simulations, one of the most important choices to be made in advance is the time-step size. Integrators can often be classified by their time-stepping implementation. Although fixed-size integrators such as the leapfrog scheme \cite{yoshida1990construction} are easier to implement, they lack the versatility of variable time-stepping methods \cite{2003gmbp.book.....H,Aarseth2003}. For example, \cite{hut1995building} presents a method to guarantee   {time} symmetry in any integration scheme when applying variable time-step sizes and emphasizes the relevance of adapting the time-step size in dynamical computer simulations of particles. Examples of integrators that implement variable step-size include {\tt Hermite} \cite{Hermite_integrator} 
integrator, which is frequently used for solving the general \textit{N}-body problem. A more optimized time-stepping scheme based on the true dynamical time of the individual particles was designed by
\cite{10.1111/j.1365-2966.2007.11427.x}.

These integrators use characteristic values of the system to estimate the optimal time step size at each integration step.   {This estimation is currently not a physical quantity but an empirical approximation}. For example, {\tt Hermite} uses the free fall time of pairs of particles as an indication of how closely the particles in the system are interacting with each other and calculates an appropriate time-step,   {as explained in} \cite{1975ARA&A..13....1A,2012NewA...17..711P}, and \cite{capuzzo2013hermite}. A more detailed description of the time-stepping implementation of {\tt Hermite} is presented in \autoref{subsec:integrator}. Even in these integrators where the time-step is determined at runtime, there is a control parameter that needs to be set manually before the start of the simulation. This so-called \textit{time-step parameter} generally has a value between
$10^{-4}$ and $1$ and scales the internally determined integration time step. The wide range of this free parameter illustrates the complexity of making an expert-knowledge-based decision. Modern integration algorithms are geared to high efficiency while conserving energy. The automated method in which the time step is decided internally can perform excellently for the right choice of the time-step parameter. A poor choice will result in energy errors that are above the acceptable values, or unnecessarily computationally expensive calculations.   {A value of $10^{-2}$ is commonly used without a convergence study that determines its optimal value}.

In traditional $N$-body calculations, the time-step parameter is fixed. While running over a relaxation timescale (or longer), the system's topology changes, rendering the pre-determined time step parameter gradually less optimal. This results in degradation of the simulation results and loss of efficiency over time. Ideally, the time-step parameter should depend on the characteristics of the system, allowing it to  {be reduced} during periods of high density, and  {increased} in episodes of low density.

Despite the problem of deciding the time-step parameter being general, it is most notorious during 3-body interactions.  We therefore  {use as an example application of our methodology a system}
of three equal-mass bodies that move under the influence of their mutual gravity.  In the 3-body problem, long-lasting wide encounters (hierarchical triples) are intermixed with short-duration close encounters (democratic triples or scrambles) \cite{2019Natur.576..406S}.

The rapidly changing conditions imply that different step sizes are needed at different points of the evolution. We overcome these drawbacks by using RL techniques for the automation of the choice of the time-step parameter during the simulation. We train the algorithm to find the optimum balance between
 accuracy and computation time.

 Although this application of RL has not yet been explored in astrophysics, \cite{dellnitz2023efficient} used reinforcement learning to choose the optimum time step for the integration of systems with rapidly changing dynamics. They overcome the inefficiencies that may arise in complex systems such as chaotic systems by combining ODE solvers with data-driven time-stepping controllers. Their scheme outperforms recently developed numerical procedures in problems in which traditional schemes show inefficient behavior. 
 
 Most examples of RL in astronomy focus on the optimal control of telescopes \cite{nousiainen2022toward, jia2023observation, telescopeRL} but RL has also been used for other applications. \cite{yi2023application} explore the use of DeepQLearning for the prediction of major solar flares and find that the performance achieved is noticeably better than that of convolutional neural networks with a similar structure. \cite{moster2021galaxynet} uses an adaptation of reinforcement learning to make predictions without unlabeled data using a reward function. 
 
 A field that presents certain similarities to astrophysics is computational fluid dynamics (CFD). In this case, the use of reinforcement learning is still at an early stage but experiencing a significant growth \cite{viquerat2022review}. Here, reinforcement learning can be used for applications such as drag reduction, shape optimization, and  conjugate heat transfer. More related to our case are applications in which reinforcement learning is used to choose simulation parameters. For example, \cite{novati2021automating} uses RL techniques to retrieve unknown coefficients in turbulence models. This field experiences challenges that can also be extrapolated to the case of astrophysical simulations. The computational cost of fluid mechanics simulations is generally high, which becomes a critical factor for the use of RL algorithms since classical methods suffer from
 low sample efficiency, i.e, they require large sample numbers to generate accurate results. Additionally, turbulence in most CFD simulations can possibly lead to a degree of stochasticity that hampers efficiency the training process. 
 A similar application is found in robotics \cite{krothapalli2011mobile}, in which reinforcement learning is used to decide the grid size (resolution) for robot navigation. The method aims to obtain finer grids next to obstacles for better resolution.

Using reinforcement learning for dynamically choosing simulation parameters in Astrophysics presents unique challenges. Fields such as fluid mechanics or weather prediction suffer from chaotic dynamics. Similarly in astrophysics, this can be perceived by the algorithm as stochasticity, which can make the training process less efficient. Additionally, computation time becomes a crucial consideration for simulations in which the number of bodies is large or a high accuracy is required. Other challenges, such as the large variability in the accelerations \cite{ulibarrena2024hybrid} and the hierarchical structure of some systems can be mitigated with simplified examples like the equal-mass 3-body problem. 
	

We study the use of DeepQLearning, a reinforcement  {learning} technique, to learn the optimal choice of time-step parameter at each step of the integration of the gravitational 3-body problem. In \autoref{sec:methodology} we present our study case and our method. In \autoref{sec:results} we present the results of our experiment and show a comparison with modern methods. In \autoref{section:symple}, we show an example of other possible uses of our method by applying it to a fixed-size integrator. 
 {The code is publicly available at} \url{https://github.com/veronicasaz/ThreeBodyProblem_astronomy}.
\section{Methodology}
\label{sec:methodology}

We present the method for integrating the motion of celestial bodies, the adopted initial conditions, our adopted DeepQLearning strategy, and the reward function.

\subsection{Integration of \textit{N}-body systems}
\label{subsec:integrator}
The gravitational force exerted on a body \textit{j} in a system of \textit{N} point masses interacting via their Newtonian gravitational forces is given by Newton's equation of motion 
\begin{equation}
\label{eq:newton}
    m_j\dfrac{d^2\vec q_j}{dt^2} = \sum_{k=0,\;
k\neq j}^{N-1}G\dfrac{m_jm_k}{\vert \vert \vec q_{jk}\vert \vert^3} \vec q_{jk}, \qquad \vec q_{jk} = \vec q_k - \vec q_j.
\end{equation}
Here ($m_j$) is the mass of particle $j$, $\vec q$ is the position vector, and $G$ is the universal gravitational constant. The indices $j$ and $k$ represent the celestial bodies. 

To calculate the evolution of a system of \textit{N}-bodies, we use a numerical integrator. For our experiment, we adopt the fourth-order accurate predictor-corrector Hermite scheme \cite{Hermite_integrator}. This integrator has a variable and individual block time-steps \cite{2008NewA...13..498N}. Variable refers to the possibility of having different values of the time-step at different steps of the simulation and individual block refers to the clustering of time steps separated by factors of two and shared among the particles in the group.  These time steps are not time-symmetric, but this can be  {approximately-}solved with a small adaptation, \cite{2006NewA...12..124M}.

We adopt the implementation for Hermite integrator from the Astrophysical Multipurpose Software Environment (AMUSE) \cite{2009NewA...14..369P,2018amuse}. The code is called {\tt Hermite}, and its time step is calculated from the free-fall time of the individual particles \cite{hut1995building}. This variable represents the time it would take two particles to collapse due to their own gravitational attraction. The time-step is calculated  according to \cite{aarseth1985direct,2008NewA...13..498N} in \textit{N}-body units as a function of the accelerations $\mathbf{a}$ and its $k$ derivatives ($\mathbf{a}^{(k)}$) as
 \begin{equation}
   \Delta t_i = \mu \left( \dfrac{\vert \mathbf {a}\vert \vert \mathbf{a}^{(2)}\vert + \vert \mathbf{\dot {\mathbf{a}}}\vert^2}{\vert \dot {\mathbf {a}}\vert \vert \mathbf{a}^{(3)}\vert+ \vert \mathbf{a}^{(2)}\vert^2} \right)^{1/2}.
   \label{Eq:timestepsize}
 \end{equation}

Here $\Delta t_i \propto \mu$, where $\mu$ is the time-step parameter, and we adopt Einstein's convention for derivatives. $\mu$ is a design parameter that can be used to increase or decrease the accuracy of the simulation. A small value of $\mu$ leads to a small time-step, and therefore to higher accuracy but longer integration time. One criterion for our approach is to optimize the total integration time.

The other criterion is the validity of the calculation. This abstract measure, in  {some cases of a} chaotic system, can be estimated using the energy error of the system. Any integrator for a problem that lacks an analytic solution introduces an energy error.  {Using this metric as an indication of accuracy is limited to cases in which all bodies have similar masses.} In \autoref{Conclusions}  {we provide further discussion on the limitations and possible future improvements for this metric}.  We calculate the energy error by taking the difference between the total energies of the system initially and at time step $i$. The total energy $E$ can be calculated as the sum of the kinetic $E_k$ and the potential energy $E_p$ of the system, and the energy error becomes
\begin{equation}
\label{eq:energy}
    \Delta E_i = \dfrac{(E_{k,i} + E_{p, i}) - (E_{k,0} + E_{p, 0} )}{E_{k,0} + E_{p, 0}} = \dfrac{E_i-E_0}{E_0}.
\end{equation}
 {A large energy error is an indication} of unphysical solutions, and the aim is to keep this error as small as possible.


\subsection{Initial conditions: the 3-body problem}
\label{subsec:initial}

The gravitational 3-body problem is the lowest-\textit{N} chaotic Newtonian dynamical system for celestial mechanics. In addition, any system of three bodies will eventually lead to the ejection of one body, leaving the other two bound in a pair \cite{1975MNRAS.173..729H}. Due to the combination of finite lifetime, chaotic behavior, and rapidly changing topologies, the gravitational 3-body problem represents an ideal study case to test and validate our RL strategy.

Since systems of three or more bodies are dynamically unstable and notoriously chaotic, any energy error inevitably leads to a different outcome.  This was demonstrated in \cite{2015ComAC...2....2B}, where they adopted the arbitrary precise \textit{N}-body code Brutus, and argued that a relative energy conservation of $10^{-4}$ suffices to preserve the physics. We adopt the same criterion here.

A proper choice of the time step is paramount for a reliable integration, small energy error, and acceptable runtime. Close encounters, which are inevitable, ensure that the integration time step has to vary over several orders of magnitude in order to comply with our main objective: find the largest possible time step for an acceptable energy error.

The 3-body problem is characterized by the masses, positions, and velocities of the three particles at some moment in time. 

For clarity, we choose the masses to be identical and equal to one solar mass ($1 \; M_{\odot}$) and we limit ourselves to two dimensions ($x$ and $y$). One particle is initialized in the center of the reference frame of the three bodies with zero velocity. The two other particles are positioned randomly following a uniform distribution with a fixed velocity according to the limits dictated in \autoref{table:inititalconditions}. We keep astronomical units (au) as a baseline for position, $M_\odot$ for mass, and $km/s$ for velocity.

\begin{table}[t]
\caption{Initial values of mass, position, and velocity for the three particles.}
\label{table:inititalconditions}
\vskip 0.15in
\begin{center}
\begin{small}
\begin{sc}
\begin{tabular}{ccccc}
\hline
Particle & $q_x \; (au)$ & $q_y\; (au)$  & $v_x\; (km/s)$& $v_y\; (km/s)$ \\
\hline
1    & 0 & 0 & 0 & 10 \\
 2 & [5, 20] & [0, 10] & -10 & 0 \\
 3    & [-10, 0] & [5, 20] & 0 & 0\\
\hline
\end{tabular}
\end{sc}
\end{small}
\end{center}
\vskip -0.1in
\end{table}

\begin{figure}[h!]
	\begin{minipage}{0.55\linewidth}
		\begin{center}
			\centerline{\includegraphics[width=\columnwidth]{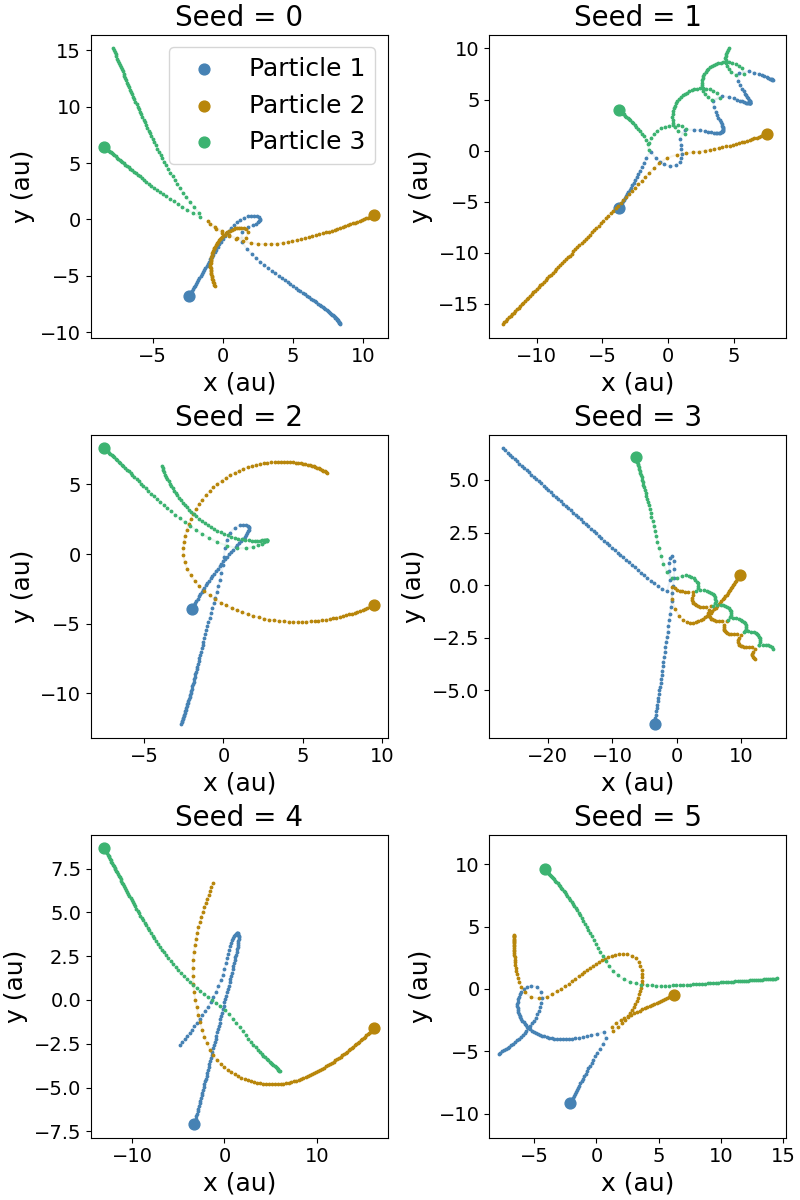}}
		\end{center}
	\end{minipage}\hfill
	\begin{minipage}{0.4\linewidth}
		\renewcommand{\arraystretch}{1.28} 
		\begin{tabular}{|ccc|}
			\hline
			Particle & $q_x \; (au)$ & $q_y\; (au)$  \\
			\hline
			\textbf{Seed 0} & &\\
			2 & 13.232 &  7.152\\
			3    & -6.028 & 13.173\\
			\hline
			\textbf{Seed 1} & &\\
			2 & 11.255 & 7.203 \\
			3    & -1.14$\times 10^{-3}$ & 9.535\\
			\hline
			\textbf{Seed 2} & &\\
			2 & 11.540 & 0.259 \\
			3    & -5.497 & 11.530\\
			\hline
			\textbf{Seed 3} & &\\
			2 & 13.261 & 7.081 \\
			3    & -2.909 & 12.662\\
			\hline
			\textbf{Seed 4} & &\\
			2 & 19.505 & 5.472  \\
			3    & -9.727 & 15.722\\
			\hline
			\textbf{Seed 5} & &\\
			2 & 8.330 & 8.707  \\
			3    & -2.067 & 18.779\\
			\hline 
		\end{tabular}
		\vspace{10pt}
		$$$$
	\end{minipage}
	\caption{Different initializations run for 100 steps with a time-step parameter of $1\times 10^{-4}$. The initial state is marked with a circle. On the right, we show the initial values associated with each random seed.}
	\label{fig:initializations}
	\vskip -0.2in
\end{figure}

In \autoref{fig:initializations}, we show the evolution of six different realizations of this system for 100 steps. Each system has an initial seed, as indicated in the various panels, and the runs are performed with $\mu = 10^{-4}$. The center of mass of the system has been initially positioned at the center of the reference frame. Each of these six systems behaves fundamentally differently on the short time scales of the integration, but more importantly, close encounters are alternated with wide excursions.  After a finite time, the systems dissolve into two bound particles and a single escaping particle. This is most noticeable in the examples with seeds 1 and 3. 


\subsection{Deep QLearning}
\label{subsec:deepql}

The adopted Reinforcement Learning method interacts with the environment, in our case the integration of the 3-body system, in order to select the optimal time-step parameter at each step. To understand the challenges associated to the use of RL in astrophysics simulations, we use a method that is interpretable and does not require the choice of a large number of training parameters. We adopt Q-learning for its relative simplicity and demonstrated efficiency compared to other RL methods \cite{sutton2018reinforcement}. Additionally, it has been shown to remain effective in stochastic environments (see for example \cite{hung2016q}).

Q-learning is a method that optimizes itself to choose an action $A$ which maximizes a reward value $R$. It is a model-free method that learns through trial and error \cite{sutton2018reinforcement} an action-value function $Q$ that approximates the optimal action-value function \cite{yi2023application}. In contrast to other popular methods, Q-learning focuses on exploration of the best possible actions and generally shows faster convergence \cite{zaghbani2024comparative}.

In our problem, the action represents the choice of time-step parameter. The action space is discrete of size  {10}, where action 0 corresponds to a time-step parameter of $\mu = 10^{-4}$ (slowest and most precise calculation) and action  {9} to $\mu = 10^{-1}$ (fastest and least accurate). For the other actions, $\mu$ has a value logarithmically spaced between these two extremes.  Using a discrete action space means that the RL algorithm is only aware of the action number, and not of the value of the time-step parameter associated with it. This allows us to be able to expand to other integrators for which the time-step calculation is implemented differently (see \autoref{subsec:extrapolintegr}). The choice of the number of actions in this work is made to allow for a simple interpretation of the results. This number can be increased to allow for a more refined algorithm.

The observation space ($S$) corresponds to the state of the system: positions, velocities, and masses of the three particles in the system. We add the current energy error in the observation space to allow the system to also account for the current accuracy of the simulation.  {For the state vector, we convert the positions and velocities to dimensionless units with base 1 au for the distance and the sum of the masses of the bodies for the mass} \cite{1971nbodyunits}.  Although a different choice of the observation space could lead to a better performance of the algorithm, we choose Cartesian coordinates to avoid expert bias in the algorithm and leave the finding of a better state vector for future work.

We adopt an extension of Q-learning called Deep Q-networks since our problem requires a continuous observation space to account for the state of the particles being continuous. A deep Q-network (DQN) combines reinforcement learning with deep neural networks \cite{mnih2015human} which are used to approximate the optimal value of the Q-function

\begin{equation}
    Q(S, A) = \text{max}_\pi \mathbb{E} \left[ R_t + \gamma R_{t+1} + \gamma^2 R_{t+2}+...  \right],
\end{equation} 

which is the maximum sum of the rewards multiplied by the discount value $\gamma$ at each time step $t$. This maximum is chosen following a behavior policy $\pi = P(A \vert S)$ after making an observation $S$ with an action taken $A$. To balance the trade-off between exploration and exploitation, the algorithm includes a stochastic exploration strategy called $\epsilon$-greedy, by which a random action is chosen with probability $p$ instead of the one selected by the algorithm \cite{viquerat2022review}. This value $\epsilon$ is reduced during the training to favor exploitation over exploration. 
To avoid the inherent instabilities of RL, the method uses experience replay, which stores the data in a training database and randomizes the chosen training sample to eliminate correlations in the observation sequence \cite{mnih2015human}. 

Deep Q-networks make use of a neural network instead of a Q table. In order to avoid instability and variability during training \cite{yu2018historical}, it is common to use two different networks: the QNet and the Target network. The first is updated at every step of the training algorithm, whereas the weights of the Target net are only updated with the ones of the QNet after multiple steps. The predictions to further select the best Q-value are made with the target net, allowing for further stability of the results than with only one network. Both networks have the same architecture and their only difference is the frequency at which their weights are updated. 

We implement our deep Q-learning method using \texttt{PyTorch} \cite{imambi2021pytorch} with an environment created using \texttt{Gym} \cite{1606.01540}.

\subsection{Reward function}
\label{subsec:reward}
The reinforcement learning algorithm will learn to choose the optimum strategy for each specific realization of the system. We define this optimum based on the balance between accuracy and computing time. 

\begin{figure}[ht]
	\vskip 0.2in
	\begin{center}
		\centerline{\includegraphics[width=0.6\columnwidth]{ 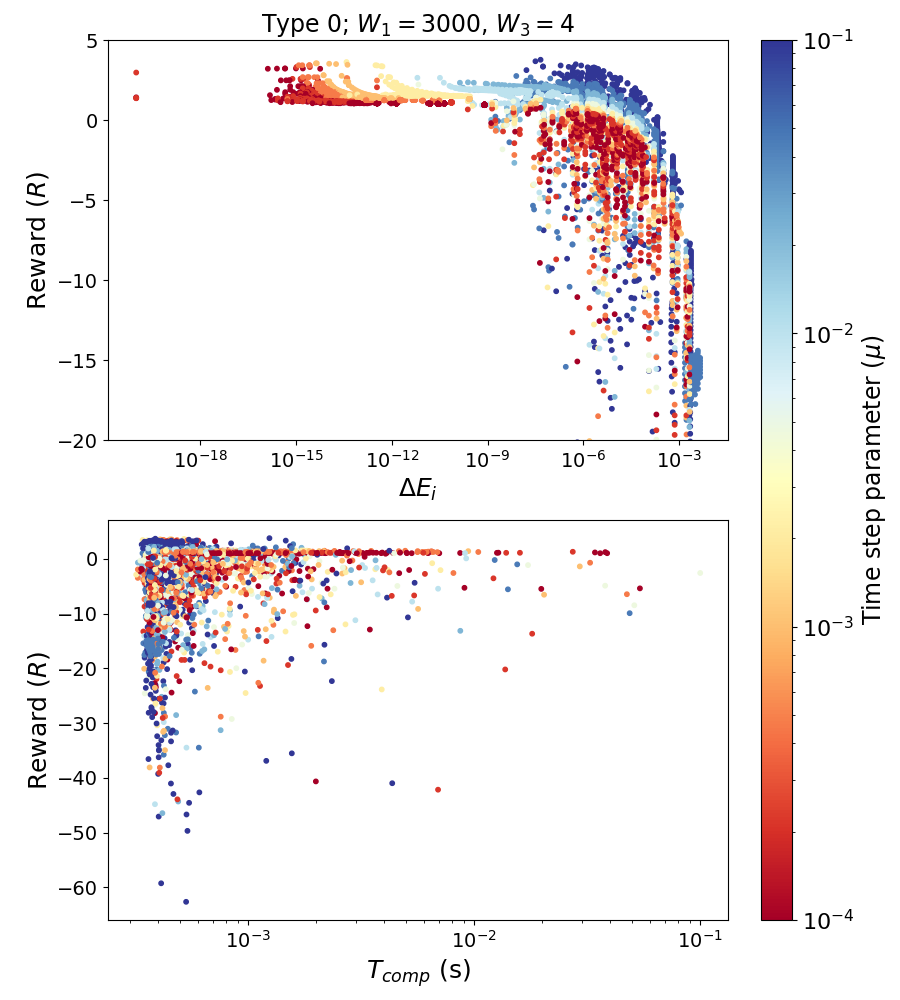}}
		\caption{Reward value as defined in \autoref{eq:reward2}. The top plot shows the reward as a function of $\Delta E_i$,  {and the bottom plot as a function of the computation time}. The color is the size of the time step parameter.}
		\label{fig:rewards}
	\end{center}
	\vskip -0.2in
\end{figure}

Our reward function consists of  {two} terms: the first accounts for the total energy error of the system at a given step, and the  {second} one for the computing time. We discuss the comparison between different expressions  {and weights} in \ref{appendix:reward}. The chosen form for the reward function is

\begin{equation} \label{eq:reward2}
	{R} = -W_0 \dfrac{1}{\text{step}}  \dfrac{\text{log}_{10} \left(\left\vert \Delta E_i \right\vert / 10^{-10} \right)}{\left\vert \text{log}_{10} (\Delta E_i) \right\vert^3} +
	 W_1 \dfrac{1}{\left\vert \text{log}_{10} (\mu)\right\vert},
\end{equation}

where $W_{[0,1]}$ represents the weights of the reward function, which are design choices. The first term in \autoref{eq:reward2} represents a  {decreasing slope with the value becoming zero} when the energy error approaches $\Delta E = 10^{-10}$,  {as can be seen in} \autoref{fig:rewards}. {We design this term so that the reward obtained by achieving energy errors below $10^{-10}$ has a smaller slope than for errors above $10^{-10}$. We achieve this by dividing the first term by the cube of the logarithm of the energy error}.  {Additionally}, we divide this term by the current integration step. By doing this, a large energy error is penalized more at the beginning of the integration than at further steps. If the energy error grows to a larger value early in the simulation, this error will continue to accumulate. By adding this term, we aim to prevent early-on large energy errors. The {second} term accounts for the computational effort by penalizing the use of a small value of $\mu$.

In \autoref{fig:rewards}, we present the value of the reward function for different steps and initializations as a function of the total energy error at that step (top panel), and as a function of  {the computation time of the step}
(bottom panel).  We observe from the first plot a decreasing trend of $R$ for larger values of the energy error. For equal values of the energy errors, a larger $\mu$ yields a larger value of $R$ to account for the computation time.  {In the bottom panel we observe that there is no clear trend for the computation time as the energy error is the dominant term.}

The weights have been chosen to balance the values of those  {two} terms and can be found in \autoref{table:trainingparams}. A comparison of different choices is also found in \ref{appendix:reward}.  {$W_0$ is chosen to be 3,000 to balance the energy error and computation time terms. This is the only weight that has to be actively chosen during the training. With this value of $W_0$, as can be seen in} \autoref{fig:rewards},  {a maximum can be achieved with either a low energy error or a large time step parameter. The value of $W_1$ is fixed to 4 for the second term to range between 1 and 4 for simplicity of the interpretation of the results. However, this value could be fixed to 1 instead, and $W_0$ would remain the only variable to be tuned. }

\subsection{Method setup}
\label{subsec:method}
\begin{figure}[h!]
\vskip 0.2in
\begin{center}
\centerline{\includegraphics[width=0.38\columnwidth]{ 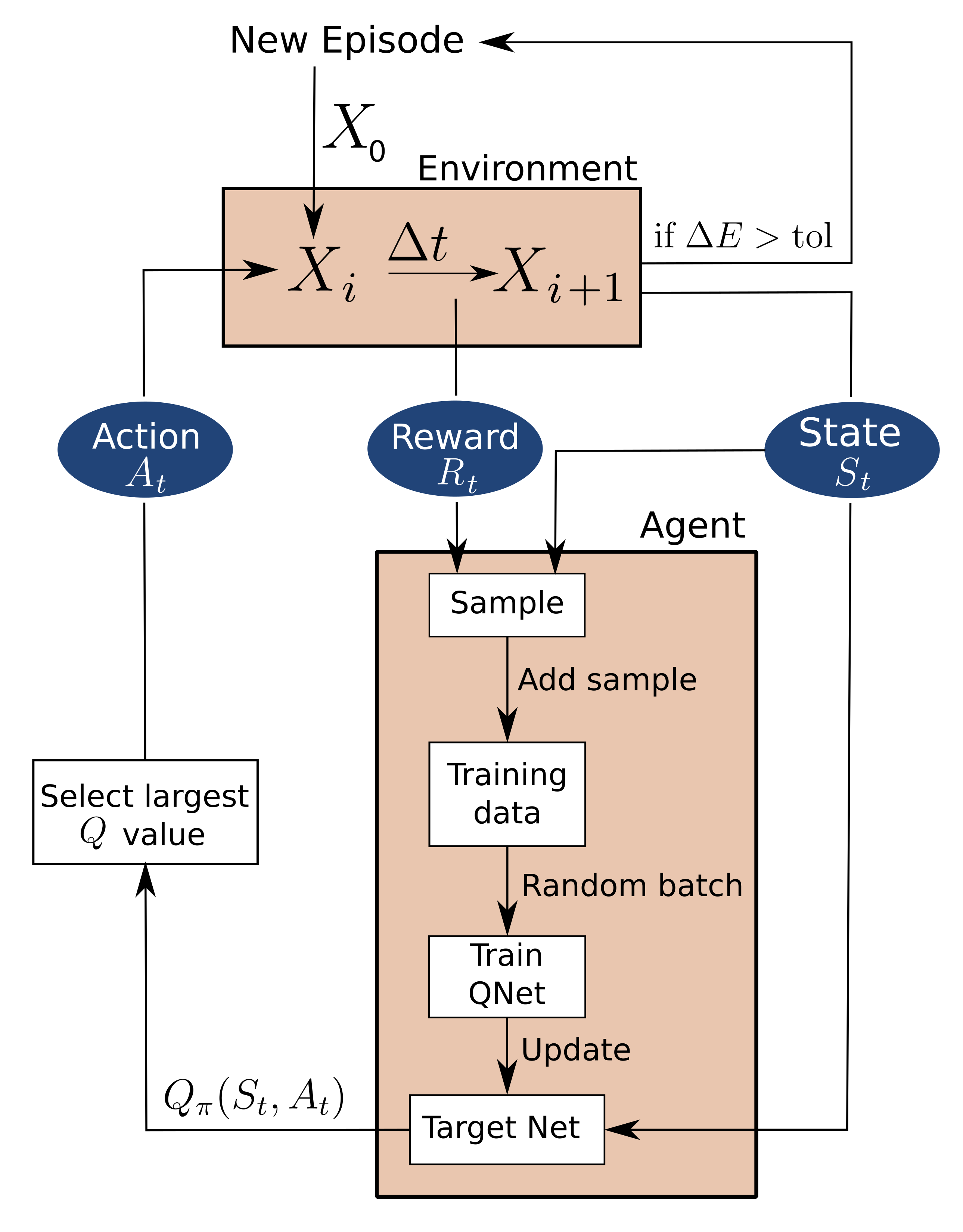}}
\caption{Schematic of the DeepQL setup. For each episode, the state of the system is evolved using the selected action, and together with the energy error and state, the QNet and the Target Networks are trained and updated, respectively, to select the optimal action.}
\label{schematic}
\end{center}
\vskip -0.2in
\end{figure}

In \autoref{schematic} we show the DQN algorithm used. Starting with an initialization of the system at time $i$ (as shown in Subsection \ref{subsec:initial}), we evolve the state ($X$) for a time ($\Delta t$), which results in the state of the system $X_{i+1}$ at time $i+1$. This means that we find the positions and velocities of the particles in the system at a later time. 
To create a training database, we take one sample from each evolution step. The sample is formed by the State ($S$) and the reward ($R$) associated with the given action ($A$) at step $t$. The reward is calculated using the energy error for that step (as in \autoref{eq:energy}) and the time-step size as explained in \autoref{subsec:reward}. The state ($S$) that is fed to the training agent is formed by the positions, velocities, and masses of the bodies ($X$) as well as the energy error (as explained in \autoref{subsec:deepql})


To train the algorithm, we generate a random batch from the training database generated and use it to update the weights of the QNet. After a given number of training steps ($t$), the Target network is updated with the weights of the QNet. The current state $S$ is used as input to the Target net to predict the Q-values associated with every possible action in the action space. Then, the action with the largest Q-value is selected to evolve the system for the next time step, and this process is repeated recursively. In some cases, a random action is taken instead of the predicted one to allow escaping local minima.

When the energy error of the simulation reaches a tolerance value of $10^{-1}$ or the maximum number of steps per episode is reached, the simulation is terminated and a new set of initial conditions is chosen to restart the process for the next episode.

\subsection{Training parameters}
\label{subsec:trainingparams}
The training parameters are selected based on manual experimentation due to the large computational resources needed for an automatized hyperparameter optimization procedure. The main hyperparameters and simulation parameters are shown in \autoref{table:trainingparams}. The first rows correspond to the maximum number of episodes, steps per episode, and the energy error tolerance to stop each episode, as explained in \autoref{subsec:method}. The latter is chosen based on the assumption that a larger energy error indicates an unphysical solution. The maximum number of steps is chosen to be 100 as it is in that interval of time when most close encounters occur, but this number could be made arbitrarily large. After a time equal to $\Delta t$, we evaluate the physical properties of the $N$-body system and choose a new action accordingly. The value of $\Delta t$ differs from the internal time-step size (see \autoref{Eq:timestepsize}). The last rows in \autoref{table:trainingparams} correspond to the number of actions chosen to train the algorithm, the minimum and maximum values of the time-step parameter $\mu$ (\autoref{subsec:deepql}), and the weights of the reward function (\autoref{eq:reward2}). 

\begin{table}[t]
\caption{Training and simulation parameters.}
\label{table:trainingparams}
\vskip 0.15in
\begin{center}
\begin{small}
\begin{sc}
\begin{tabular}{lc}
\textbf{Global search}&\\
\hline
Max episodes & 2,000 \\
Max steps per episode & 100 \\
$\Delta E$ tolerance & $1\times 10^{-1}$\\
$\Delta t$ & $1\times 10^{-1}$ (yr) \\
\hline
Hidden layers & 9\\
Neurons per layer & 200\\
Learning rate & $1\times 10^{-3}$\\
Batch size & 128 \\
Test data size & 5\\
\hline
Number of actions & 10 \\
$[\mu_{\text{min}}, \;\mu_{\text{max}}]$ & $[1\times 10^{-4}, \;1\times 10^{-1}]$\\
$W_{[0,1]}$ & $[3,000 \;,\; 4.0]$\\
\hline
\end{tabular}
\end{sc}
\end{small}
\end{center}
\vskip -0.1in
\end{table}

\begin{figure}[ht]
	\vskip 0.2in
	\begin{center}
		\centerline{\includegraphics[width=0.6\columnwidth]{ 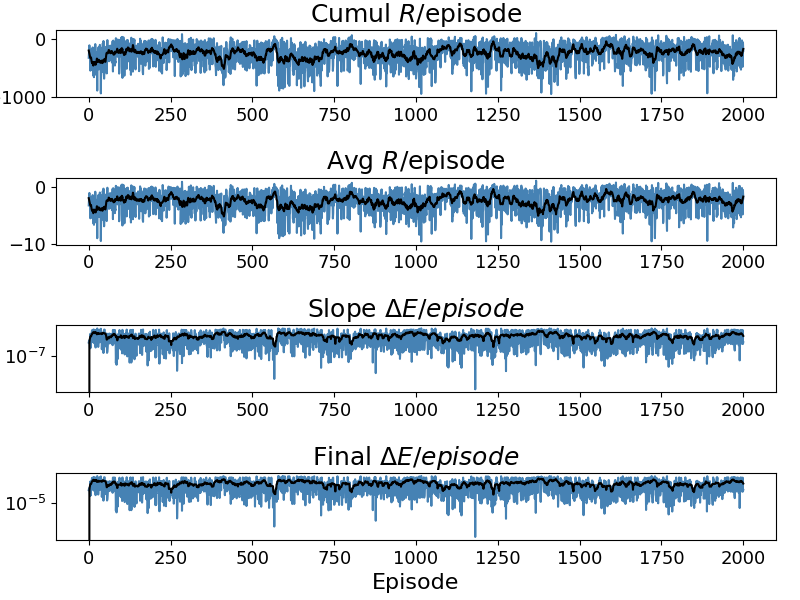}}
		\caption{Evolution of the cumulative reward per episode (first row), the average reward per episode (second row), the rate of growth of the energy error per episode (third row), and the final energy error per episode (fourth row).}
		\label{cumulative_reward}
	\end{center}
	\vskip -0.2in
\end{figure}

 In \autoref{cumulative_reward}, we show the evolution of the training process at each episode using the cumulative reward, the average reward, the slope of the energy error, and the final energy error. We can observe large oscillations due to the presence of random functions in the training procedure and the large differences in behavior between the various initializations. We show in black a fit of these oscillations. However, due to the chaotic nature of the problem, it is not possible to use these metrics as a measurement of the quality of the training procedure. 

\begin{figure}[h]
	\centering
	\begin{subfigure}[b]{0.49\textwidth}
		\includegraphics[width=\textwidth]{ 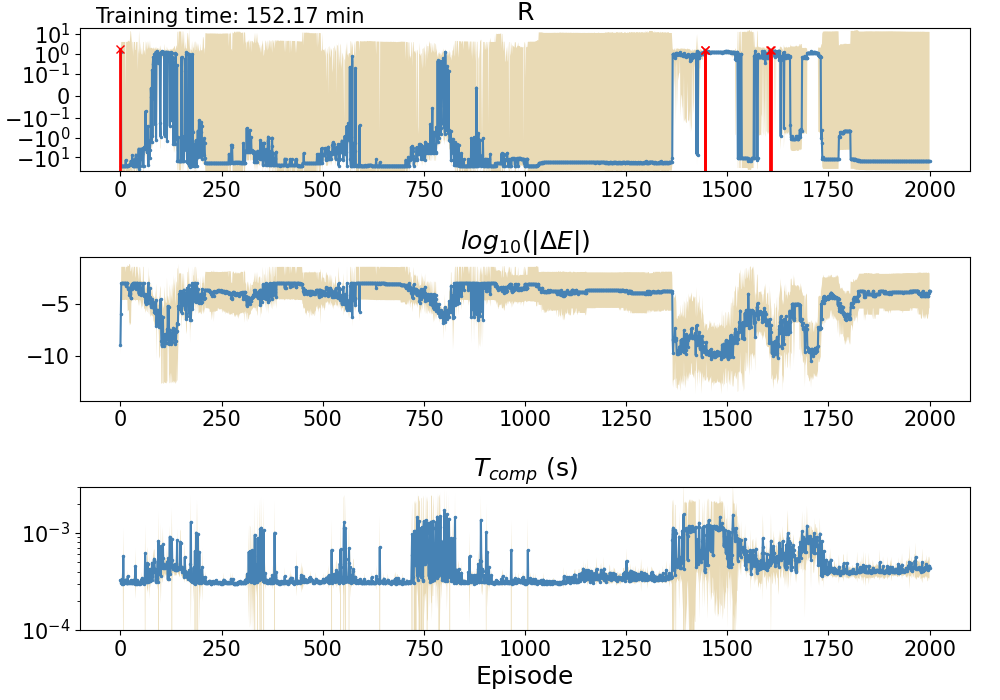}
		\caption{}
		\label{fig:test_reward_1}
	\end{subfigure}
	\begin{subfigure}[b]{0.49\textwidth}
		\includegraphics[width=\textwidth]{ 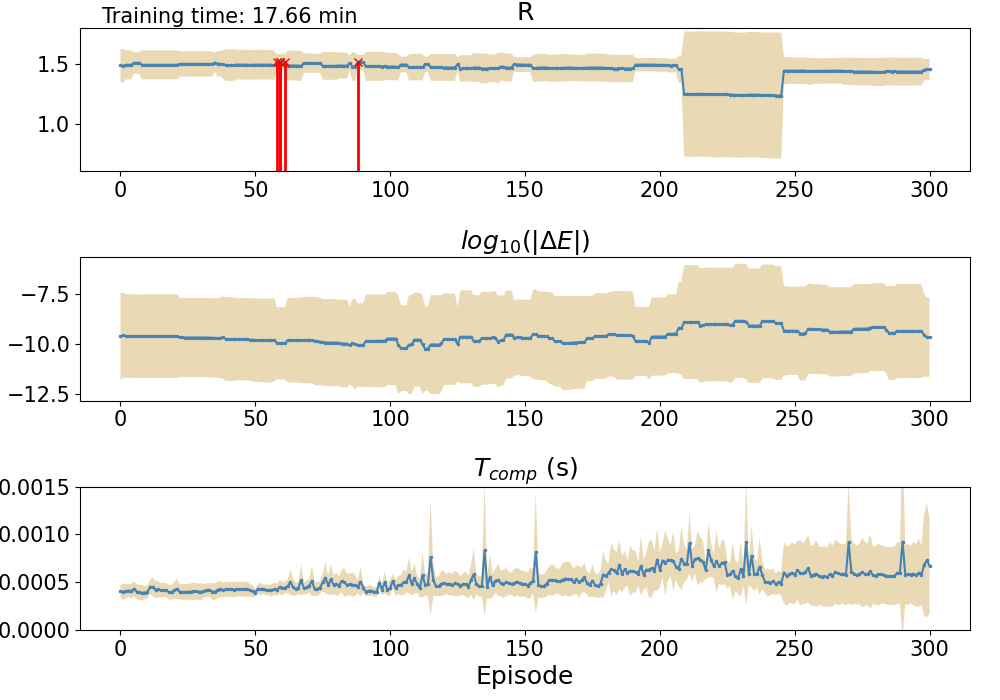}
		\caption{}
		\label{fig:test_reward_12}
	\end{subfigure}
	
	\caption{Evolution of the average  {(blue)} and standard deviation  {(orange)} of different metrics of the test dataset per\textit{} episode of: the reward value (first row), the energy error (second row), and the computation time (third row) for the global training \textbf{(a)} and for the local training performed after the global one \textbf{(b)}.}
\end{figure}

To overcome this problem, we create a test dataset that contains a set of initial conditions. At the end of each training episode the performance of the RL algorithm is tested on those initializations and the results are saved as a measurement of the evolution of the training procedure. In this way, we prevent randomness in our evaluation of the training by using the same initial realizations at each episode and therewith create a robust testing method. It is important to mention that this method incurs in additional computation time. Depending on the needs of the experiment, the testing can be done every X episodes to save computational effort. The size of the test data used is specified in \autoref{table:trainingparams}. In \autoref{fig:test_reward_1}, we show the same training as in \autoref{cumulative_reward} but with our testing method. We plot the average and standard deviation of each of the metrics for the testing samples. It is now possible to distinguish between the models with high reward and those  {which} perform poorly. We then mark with a red line those episodes in which the reward achieved is the largest. We explore those models and others with similar reward values and choose the model at episode  {1444} as our best-performing candidate. 

 {The training time changes depending on the actions taken at each episode and the test data size. With our settings, training for 300 epochs takes 17 min (on an AMD Ryzen 9 5900hs). The training times are included in the figure corresponding to the training evolution} (\autoref{fig:test_reward_1}).  

The reward function landscape is made of many local maximums. Since this function is composed of two terms, a local maximum could be one of the solutions in which the algorithm always chooses the most restrictive actions. In this case, the energy error term would be maximized at the cost of computation time. On the other hand, another local maximum could be the choice of the least restrictive action to maximize the computation time term. We aim to obtain an algorithm that can balance those two, and it is therefore essential to choose the correct weights for the reward function. Even when the weights are optimal, the random nature of the training may lead the algorithm towards one of these local maximums. We therefore repeat the training multiple times with different seeds until the maximum reward represents our objectives.  {The model chosen comes from a training with an initial weight distribution corresponding to seed 1.}

The model chosen is capable of identifying certain features of the integration and adapting to it. However, we did not find the result fully satisfactory. Therefore, we perform a local search starting from the weights of the selected model. Using the training conditions shown in \autoref{table:trainingparams2}, we further train the RL algorithm for  {300} episodes with a lower learning rate to find solutions that are close in the optimization landscape to the one obtained with the global search. This training is shown in \autoref{fig:test_reward_12}. We evaluate the models at the episodes with the largest reward value and take the best-performing one. We choose the model at episode  {88} for our final results. This procedure could be further repeated to obtain consecutively better results, but on account of the computation time limitation, we  {restrict} ourselves to one local search.

\begin{table}[t]
	\caption{Training and simulation parameters of the local search}
	\label{table:trainingparams2}
	\vskip 0.15in
	\begin{center}
		\begin{small}
			\begin{sc}
				\begin{tabular}{lc}
					\textbf{Local search}&\\
					\hline
					Max episodes & 300 \\
					Learning rate & $1\times 10^{-8}$\\
					Test data size & 5\\
					\hline
				\end{tabular}
			\end{sc}
		\end{small}
	\end{center}
	\vskip -0.1in
\end{table}

\section{Results}
\label{sec:results}

In this section, we discuss the results of the integration of a system of three particles using the trained RL algorithm and compare it with the canonical approach  {without retraining}. 

\subsection{Integration results}
 
 {Once the algorithm has been trained using the parameters shown in} \autoref{subsec:trainingparams}  {the model can be used repeatedly without the need for retraining or without any expert knowledge involved.} We aim to obtain an algorithm that chooses small values of the time-step parameter when close encounters occur, and large values when the particles are farther away from each other. One remaining challenge is the gradually growing energy error, and the fact that once increased, is it not likely to decrease by large amounts. We mitigate this problem by including the energy error in the state $S$ of the DQN algorithm. The algorithm then takes the relative local error and the global energy error into account for making a decision.

\begin{figure}[h]
	\centering
	\begin{subfigure}[b]{0.49\textwidth}
	\includegraphics[width=\textwidth]{ 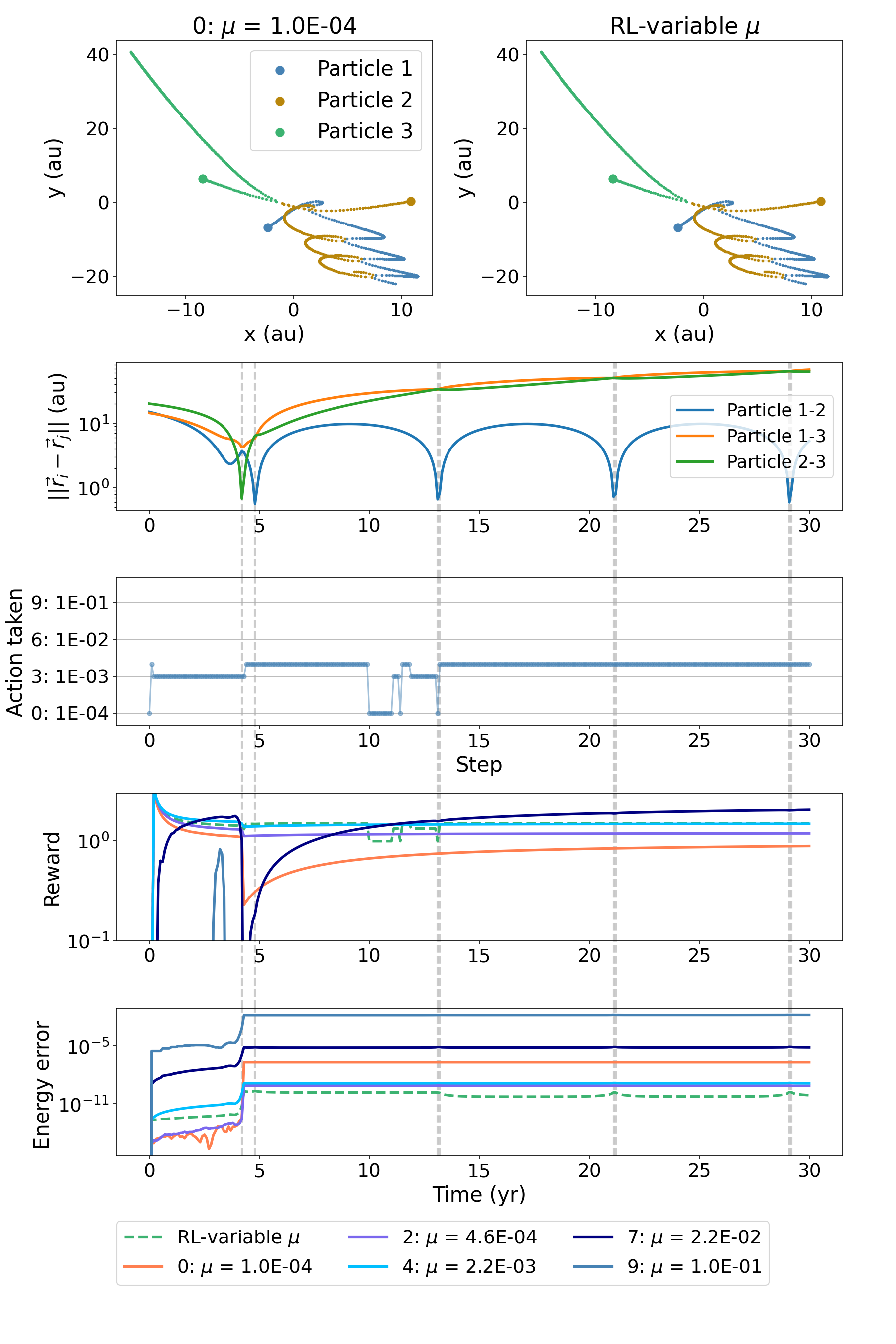}
	\caption{}
	\label{fig:comparisonRLvsfixedsize1}
	\end{subfigure}
	\begin{subfigure}[b]{0.49\textwidth}
		\includegraphics[width=\textwidth]{ 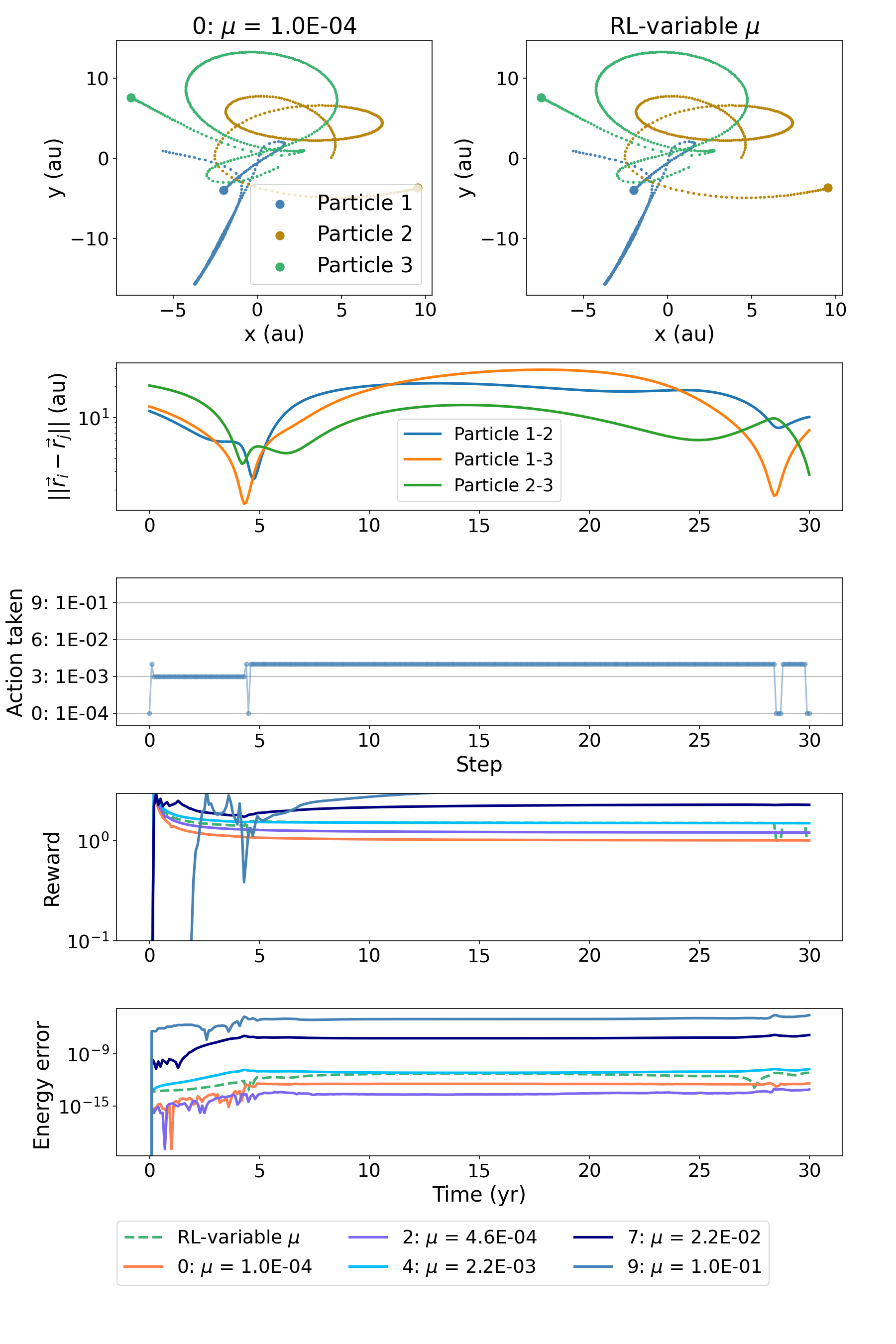}
		\caption{}
		\label{fig:comparisonRLvsfixedsize2}
	\end{subfigure}
	
	\caption{Comparison of fixed-size time-step parameters with our RL model for 300 time steps (30 years). We present the trajectory in Cartesian coordinates (top row panels), the pairwise distance between particles (second row), the actions taken by the RL algorithm (third row), the reward for each case (fourth row), and the energy error at each time step for each study case (fifth row), for initializations with Seed 0 \textbf{(a)} and Seed 2 \textbf{(b)}.}
\end{figure}

\begin{figure}[h!]
	\vskip 0.2in
	\begin{center}
		\centerline{\includegraphics[width=0.6\columnwidth]{ 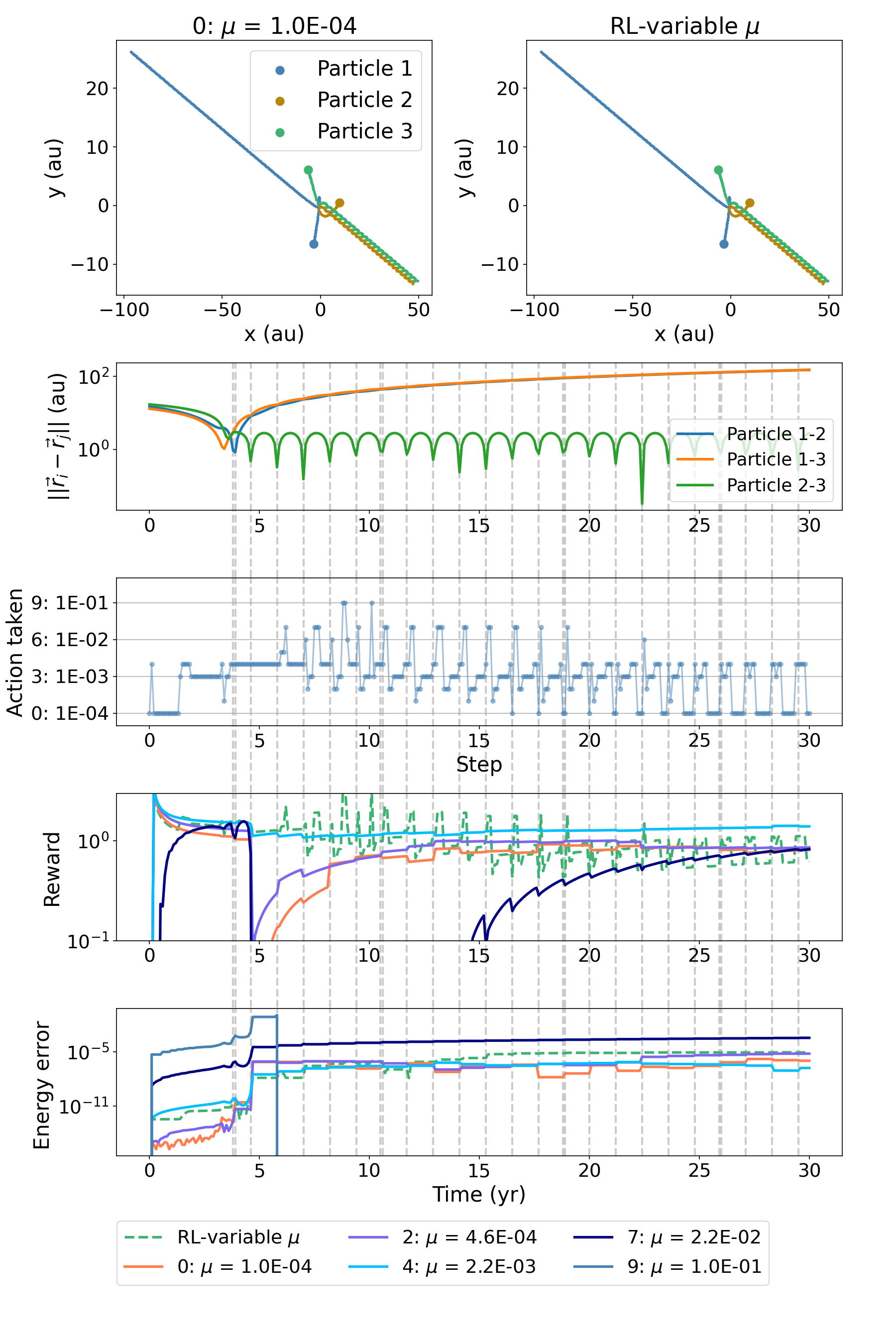}}
		\caption{ {Comparison of fixed-size time-step parameters with our RL model for the initialization with Seed 3 run for 300 time steps (30 years). We present the trajectory in Cartesian coordinates (top row panels), the pairwise distance between particles (second row), the actions taken by the RL algorithm (third row), the reward for each case (fourth row), and the energy error at each time step for each study case (fifth row).}}
		\label{comparisonRLvsfixedsize7}
	\end{center}
	\vskip -0.2in
\end{figure}

To illustrate the working of our trained RL algorithm, we evolve the initial conditions (see \autoref{fig:initializations}) for 300 time steps (30 years). Since our algorithm was trained for the integration up to 100 steps, training to 300 steps allows us to also understand the extrapolation capabilities of the trained model. The results are presented in \autoref{fig:comparisonRLvsfixedsize1}, \autoref{fig:comparisonRLvsfixedsize2}, and \autoref{comparisonRLvsfixedsize7}.

In the first row, we show the trajectory of the three particles using a fixed value for $\mu = 10^{-4}$ (left), and the results of the RL algorithm (right). They look indistinguishable. In the second row, we present the pairwise mutual distance between the three particles. This is a good method to identify close encounters. The third row gives the action chosen by the RL algorithm for each time step. The last two rows represent the reward and the energy error for each study case, respectively. 

We can observe how the trained RL model correctly identifies close encounters and adopts a more restrictive action to prevent a large jump in energy error  {when necessary}. When the distance between the particles  {does not correspond to close encounters}, our trained model increases the time-step parameter, leading to faster calculations with acceptable accuracy. The selected action is smaller during close encounters and larger when the system is hierarchical. 

Our initial intuition led us to think that the optimal RL algorithm would select a less restrictive action unless there is a close encounter, in which case a more restrictive action is needed. Therefore, we expected lower peaks in the action plot (third row of \autoref{fig:comparisonRLvsfixedsize1}) when the bodies have a close encounter. However, studying our trained RL model, we observed that the high-reward solutions many times skip some of those peaks  {without an increase in the energy error}. If we take into consideration that the energy error will not decrease substantially as the simulation progresses, we realize that contrary to our initial intuition, choosing a smaller time-step parameter might not be the optimal choice once our simulation has advanced. As an example, if at step 50 our energy error is on the order of $10^{-7}$, we need to choose a less restrictive action during the close encounter to keep the energy error lower than that. However, if the energy error after 50 steps is $10^{-4}$, and the cost of choosing a less restrictive action is of the same order, choosing a restrictive action to achieve a lower energy error will not have a noticeable effect in the global simulation. 

We show the results of our trained RL algorithm for  {three representative} initializations of the 3-body problem. 
For the result with Seed  {0} in \autoref{fig:comparisonRLvsfixedsize1}, we observe how the algorithm correctly identifies close encounters and adapts by choosing a lower action. In some close encounters, this behavior is not seen. As explained before, choosing a lower action is only necessary depending on our current energy error. In some cases, choosing for a lower action will not lead to an improvement of the accuracy. This is the case, as we can observe that even by skipping a close encounter, the energy error still remains constant throughout the simulation. Additionally, this simulation is run to 300 steps (30 years), whereas the training is done with simulations up to 100 steps, which speaks of the excellent extrapolation capabilities of our method.  {In the case of Seed 2 shown in} \autoref{fig:comparisonRLvsfixedsize2}{, the close encounters are less pronounced, but the algorithm still identifies them correctly and chooses a lower time-step parameter accordingly. Finally, in the results for Seed 3} (\autoref{comparisonRLvsfixedsize7})  {we can observe a close encounter between the three bodies followed by consecutive close encounters between two of them. In this case, the algorithm adapts at each encounter to prevent the energy error from increasing. This leads to a fast oscillation of the actions taken at each step. }

\begin{figure}[ht]
\vskip 0.2in
\begin{center}
\centerline{\includegraphics[width=0.5\columnwidth]{ 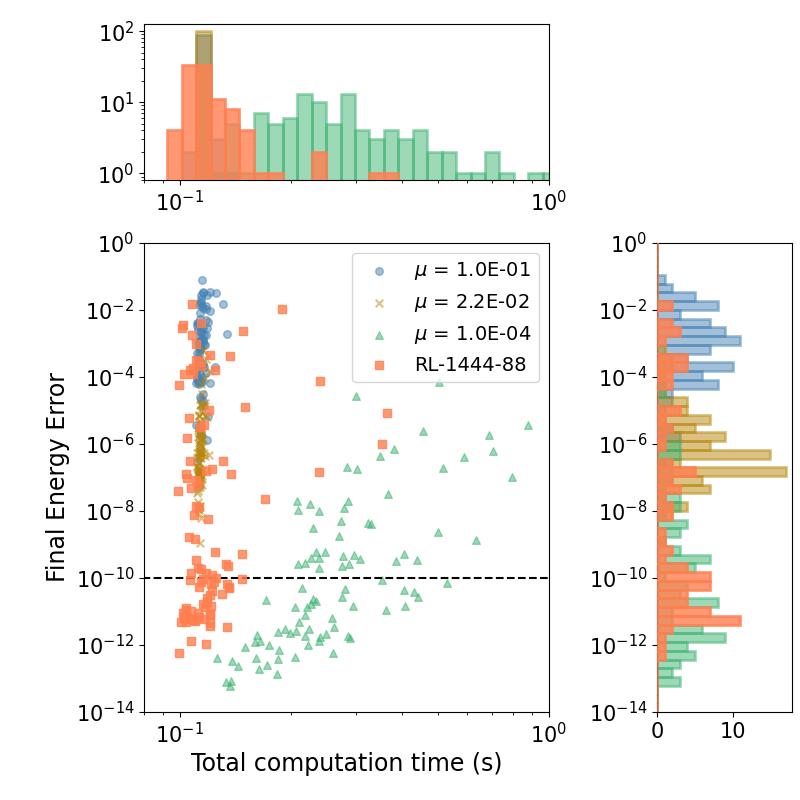}}
\caption{Distribution of computation time and energy error at the last step for 100 initializations run for 300 steps (30 years). Each color represents one study case, from the most inaccurate fixed value in  {blue} ($\mu = 10^{-1}$), to the most accurate ones ($\mu = 10^{-4}$) in green. The results for the RL algorithm using variable $\mu$ are presented in  {orange}.}
\label{fig:results_compar2}
\end{center}
\vskip -0.2in
\end{figure}

To better appreciate the performance of the trained RL algorithm, we present in \autoref{fig:results_compar2} the energy error and computation time at the end time of 100 initializations run for  {300} steps. We compare the RL algorithm ({orange}) with a rather inaccurate fixed value $\mu = 10^{-1}$ ({blue}), and with a rather accurate fixed value $\mu =10^{-4}$ ({green}). As expected, the calculations for $\mu =10^{-1}$ produce larger energy errors but require lower computation effort, whereas the calculations for $\mu = 10^{-4}$ are generally more precise at a larger computational cost. The majority of the calculations with $\mu = 10^{-4}$ are unnecessarily accurate as they conserve energy better than $10^{-10}$, but at a cost of more than twice the computation time of our RL solution.  {It should be noted that the computation time refers to the integration time, and does not include the prediction time of the network. In most problems, this time is negligible compared to the time it takes to integrate the system.}  {We additionally show the results for the case with $\mu$ closest to  $10^{-2}$ since that is the most commonly used value in astronomy simulations. Our RL algorithm achieves better energy errors than this commonly-used value while avoiding the large computation times corresponding to a low value of $\mu$. Additionally, it does so without any expert knowledge needed to set up the simulation. This result is promising, further fine-tuning of the training procedure could lead to even better results.} 

\section{Generalization capabilities}
\label{sec:extrapolation}

In this section, we discuss the generalization potential of our method when choosing a different integrator.

\subsection{Generalization capabilities: different integrators}
\label{subsec:extrapolintegr}
The RL algorithm is set up so that it can be applied to any integrator without the need to modify its source code. This means that applying it to different integrators is effortless. Therefore, we study the generalization capabilities of our method when applied to other integrators. 

The trained algorithm chooses an action ranging from 0 to  {10}, where the first is the most accurate time-step parameter and the second is the least. Some integrators like {\tt Huayno} \cite{2012NewA...17..711P} also use the time-step parameter, whereas others can use other parameters to determine the size of the time step. For example, the {\tt Symple} integrator allows choosing the order and the time step size. We use the trained algorithm for these two other integrators with the ranges of actions in \autoref{table:integrators} and compare their performance. We also adopt the implementation of these algorithms from AMUSE 
\cite{2009NewA...14..369P,2018amuse}.
\begin{table}[t]
\caption{Ranges of actions for different integrators}
\label{table:integrators}
\begin{center}
\begin{small}
\begin{sc}
\begin{tabular}{lccc}
\hline
& Value & Minimum value & Maximum value\\
\hline
Hermite & $\mu$ & $  10^{-4}$& \;$  10^{-1}$\\
Huayno & $\mu$ &$ 10^{-5}$& \;$  10^{-1}$\\
Symple & $\Delta t$ &$ 10^{-7}$& \;$  10^{-2}$\\
\hline
\end{tabular}
\end{sc}
\end{small}
\end{center}
\vskip -0.1in
\end{table}

\begin{figure}[ht]
\vskip 0.2in
\begin{center}
\centerline{\includegraphics[width=0.4\columnwidth]{ 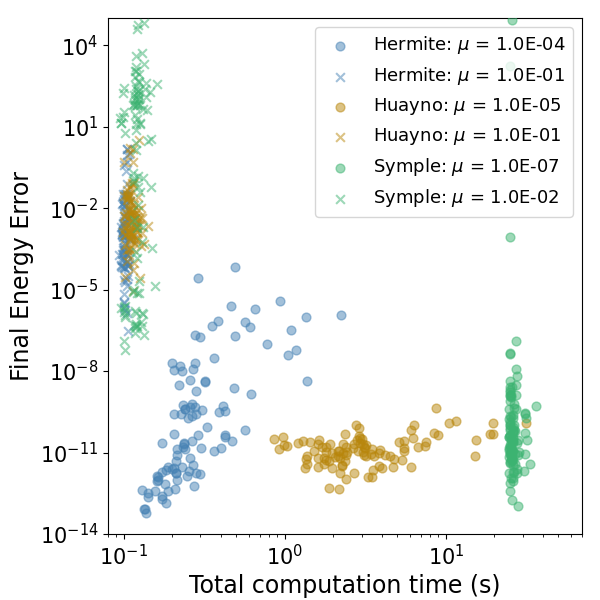}}
\caption{Comparison of the final energy error and computation time for the least and most accurate actions with {\tt{Hermite}}, {\tt{Huayno}}, and {\tt{Symple}} integrators run for 100 steps. }
\label{fig:integrator_comparsion}
\end{center}
\vskip -0.2in
\end{figure}

The ranges for the actions are chosen for the integrators to have similar values of computing time and energy error for their least and most accurate actions. We see in \autoref{fig:integrator_comparsion} the final energy error and computing times for 100 initializations integrated for 100 steps. We observe how {\tt Symple}, even with the most accurate time-step size, has a large range of final energy errors due to its fixed time-step size.  {It therefore requires a small time-step size to achieve similar energy errors as with }{\tt Hermite},  {which leads to a large computation time}.  {Similarly, we choose a smaller lower value for the time-step parameter, which leads to the simulations being slightly more computationally expensive.} 

We evolve the system with seed 0 for 300 steps (30 years) with the aforementioned integrators. In \autoref{fig:results_compar}, we show the actions taken for each integrator and the energy error incurred. Similarly to {\tt Hermite}, the RL algorithm with \texttt{Huayno} can  {identify close encounters and adopt a more restrictive action to} keep the energy error reasonably constant during close encounters, although the final energy error is approximately  {two} orders of magnitude larger. From the results obtained with {\tt Symple}, we can conclude that the trained algorithm does not extrapolate well to fixed time-step algorithms  {as it reaches the maximum allowed value of energy error early in the simulation}.

\subsection{Training and integration with Symple}
\label{section:symple}
 {In }\autoref{subsec:extrapolintegr},  {we learn that the trained model does not extrapolate well for the  \texttt{Symple} integrator. This was to be expected since \texttt{Symple} does not have an internal integration calculation and the RL method is directly choosing the time step instead of the time-step parameter. This level of extrapolation cannot be expected from current RL techniques. Although the trained model does not extrapolate, we show that our method can be extrapolated without the need for changes in the reward function or other RL parameters. To do so, we use the same method to train a new model suited for \texttt{Symple}. Similarly, this method could be applied to other integrators and problem setups. }{We train a new model with the parameters shown in} \autoref{table:trainingparams3}.  {The training metrics and the results of applying the model to the integration with }\texttt{Symple}  {can be found in} \ref{appendix:resultsSymple}.
\begin{table}[t]
	\caption{Training and simulation parameters for \texttt{Symple}}
	\label{table:trainingparams3}
	\vskip 0.15in
	\begin{center}
		\begin{small}
			\begin{sc}
				\begin{tabular}{lc}
					\textbf{Global search} &\\
					\hline
					Max episodes & 3000 \\
					Learning rate & $1\times 10^{-3}$\\
					Test data size & 5\\
					\hline
					Number of actions & 10 \\
					$[\mu_{\text{min}}, \;\mu_{\text{max}}]$ & $[1\times 10^{-7}, \;1\times 10^{-2}]$\\
					$W_{[0,1,2]}$ & $[3,000\;,\; 4]$\\
					\hline
				\end{tabular}
			\end{sc}
		\end{small}
	\end{center}
	\vskip -0.1in
\end{table}

\begin{figure}
\begin{subfigure}[t]{0.49\textwidth}
	\centerline{\includegraphics[width=\columnwidth]{ 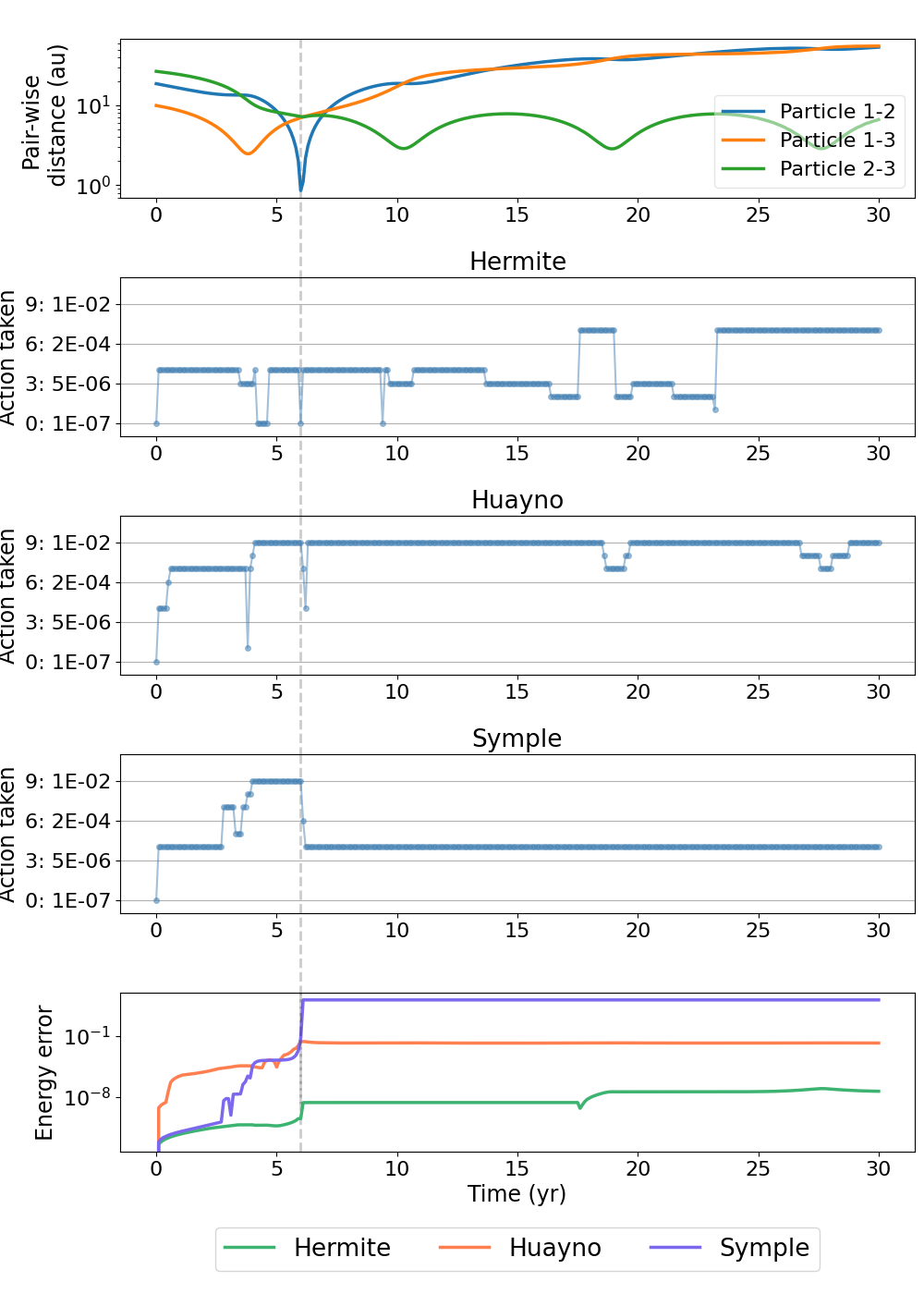}}
	\caption{}
	\label{fig:results_compar}
\end{subfigure}
\begin{subfigure}[t]{0.49\textwidth}
		\centerline{\includegraphics[width=\columnwidth]{ 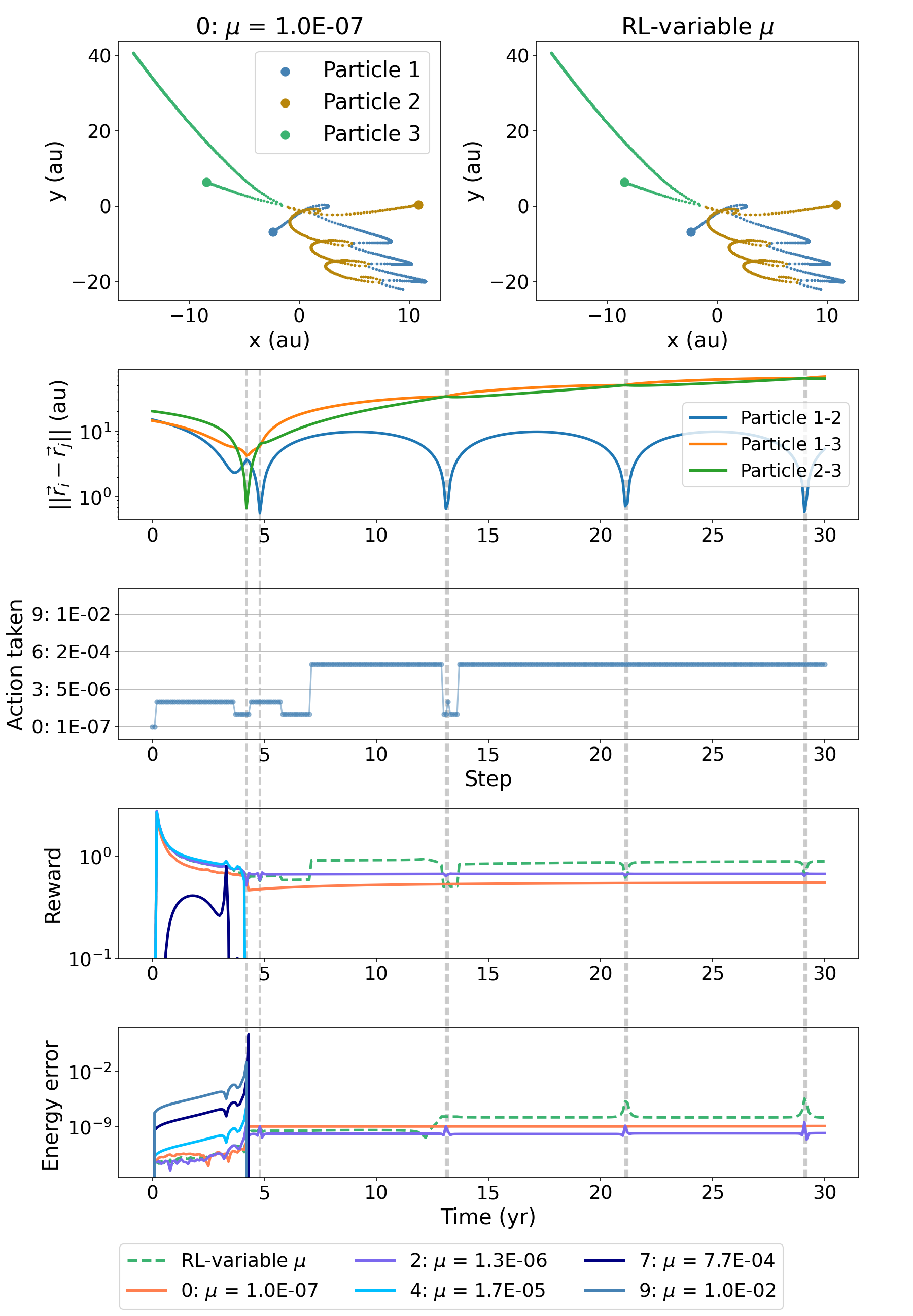}}
	\caption{}
	\label{fig:symplesteps}
\end{subfigure}
\caption{\textbf{(a)} Comparison of actions taken and energy error for three different integrators. The top panel shows the pairwise distance between bodies. \textbf{(b)} Comparison of fixed-size time step with our RL model for the initialization with Seed 0 for 300 time steps (30 years). We present the trajectory in Cartesian coordinates (top row panels), the pairwise distance between particles (second row), the actions taken by the RL algorithm (third row), the reward for each case (fourth row), and the energy error at each time step for each study case (fifth row).}
\end{figure}

We apply the trained model to the evolution of the initialization with seed 0 for  {30 years}. The results in \autoref{fig:symplesteps} can be compared with those in \autoref{fig:results_compar}. In contrast with the model trained for \texttt{Hermite}, this new model can correctly identify close encounters and adapt the actions taken accordingly. Whereas the least restrictive cases with fixed time step reach the maximum allowed energy error early in the integration, our trained model is able to adapt the actions to finish the integration with a  {low} energy error.

\begin{figure}[ht]
	\vskip 0.2in
	\begin{center}
		\centerline{\includegraphics[width=0.5\columnwidth]{ 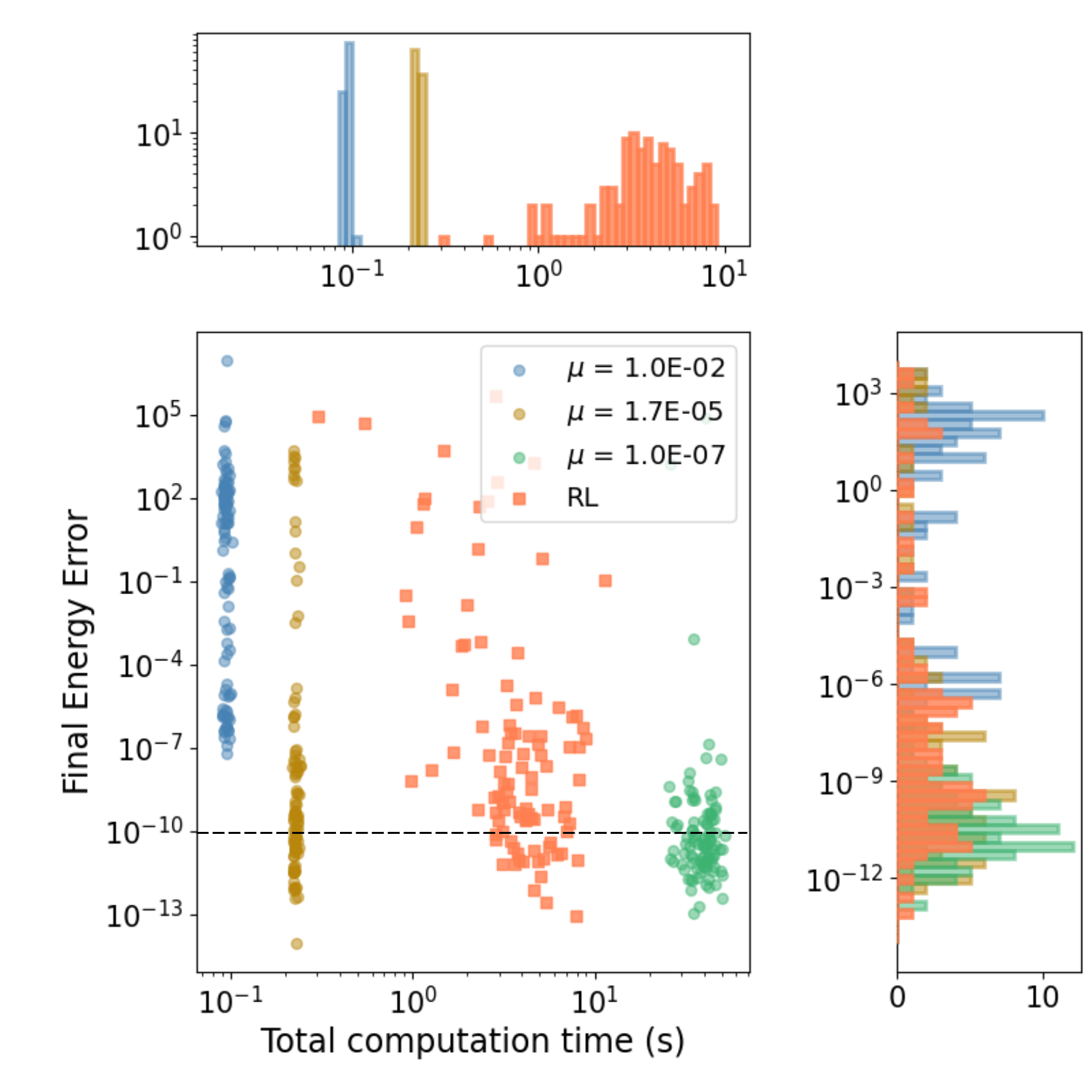}}
		\caption{Distribution of computation time and energy error at the last step for  {100} initializations run for 300 steps. Each color represents one study case, from the most inaccurate fixed value in blue, to the most accurate one in green. The results for the RL algorithm using variable $\mu$ are presented in orange.}
		\label{fig:results_compar3}
	\end{center}
	\vskip -0.2in
\end{figure}

Finally, we show the distribution of computation time and energy error for 100 initializations run for 300 steps for \texttt{Symple} in \autoref{fig:results_compar3}. The main problem with using \texttt{Symple} integrator is the large spread in final energy error for different initializations. By not having an adaptable time-step the accuracy of the simulation will depend on how chaotic each scenario is.  {We show that our RL model can achieve results that are mostly concentrated around low energy values, but the computation times achieved are not good enough to consider the trained model to be competitive. We believe that further training with optimized training parameters and a larger network would lead to a significant improvement of these results. We will leave that experiment for future work. }

\section{Discussion and conclusions}
\label{Conclusions}

We designed a method in which an agent is trained to choose the best time-step parameter for the simulation of the chaotic 3-body problem. We have overcome some main disadvantages of modern integrators, such as our baseline {\tt Hermite} integrator. Firstly, we eliminate the need for expert knowledge when choosing this parameter. Secondly, we allow it to change during the simulation to adapt to the rapidly changing conditions of the problem. Finally, the algorithm is trained to balance computation time and accuracy. 

 {The reward function determines the correct functioning of the method. Currently, we use the energy error as a metric of the accuracy of the simulation. Although this applies to the case shown here, when the ratios of the masses of the bodies are different, the energy error becomes an unreliable metric for the correctness of the dynamics. In this case, the energy contribution of the bodies with low mass is orders of magnitude smaller than that of larger bodies. Our method could be made applicable to the hierarchical \textit{N}-body problem by changing the energy term of the reward function with another metric} \cite{rauch1999dynamical,pham2024family}. 

We have shown our implementation of the reward function containing  {two} different terms to account for the energy error and computation time. We observe that our method is capable of achieving better energy conservation than most of the fixed time-step parameter cases for equivalent values of energy error and computation time. Our trained network can be used without the need for retraining in the case of other integrators. However, the performance achieved for fixed time-step size integrators is worse than that for variable time-step size methods. Although the final energy error for integrators other than {\tt Hermite} is larger, by retraining our model we can adapt its use to other integrators, such as {\tt Symple}, to achieve better performance with few training episodes. We therefore show how our method can be easily generalized to other integrators without the need to adapt their implementation.  Similarly, this framework can be used in many other similar problems with the necessary adaptations in reward function. Additionally, the state vector depends on the number of bodies in the system. In order to apply this method to higher-\textit{N} problems, the model would have to be retrained. Overcoming this issue is left for future work. 

Currently, our method performs the observation and chooses the action after a check time $\Delta t$. In future work, we aim to eliminate this requirement and perform the check after every internal time step. By doing this, the integration becomes more flexible and the time-step parameter can be adapted after smaller steps when close encounters occur. {Furthermore}, the hyperparameter optimization was done via manual experimentation. In our opinion, a systematic approach would yield better training results and help improve the final performance of the algorithm. Additionally, more complex RL methods could also contribute to an improvement in the method performance.

 {Although the results shown are preliminary, they open a new scientific opportunity for the inclusion of Machine Learning into astronomical simulations. The method presented is general enough to be applied to a variety of integrators and cases without major changes in the method itself and shows to be promising in eliminating the need for expert knowledge for setting up a simulation. }

\section{Acknowledgments}
This publication is funded by the Dutch Research Council (NWO) with project number OCENW.GROOT.2019.044 of the research programme NWO XL. It is part of the project ``Unraveling Neural Networks with Structure-Preserving Computing". In addition, part of this publication is funded by the Nederlandse Onderzoekschool Voor Astronomie (NOVA).


\bibliographystyle{ieeetr}
\bibliography{main}

\appendix
\section{Reward function comparison}
\label{appendix:reward}

Finding the right reward function for the reinforcement problem is vital to obtain optimal results. We try different combinations of the two terms: the first being the energy error at a certain step, and the second the time step parameter to account for the computational time. Those are shown in \autoref{eq:rewards}  {and} \autoref{eq:reward4}:

\begin{align}
	\label{eq:rewards}
	\text{Type 1: } \qquad&\qquad
	{R} = -W_0 \dfrac{\left( A+10 \right)}{\vert A\vert^3} +  W_1 C,\\
	\text{Type 2: } \qquad&\qquad
	{R} = -W_0 A +  W_1 C, \label{eq:reward4}
\end{align}

where
\begin{align}
A = \text{log}_{10} \left(\vert \Delta E_i \vert \right),\qquad
B = \dfrac{1}{\vert \text{log}_{10} (\mu)\vert}.
\end{align}

\begin{figure}[ht]
\vskip 0.2in
\begin{center}
\centerline{\includegraphics[width=0.68\columnwidth]{ 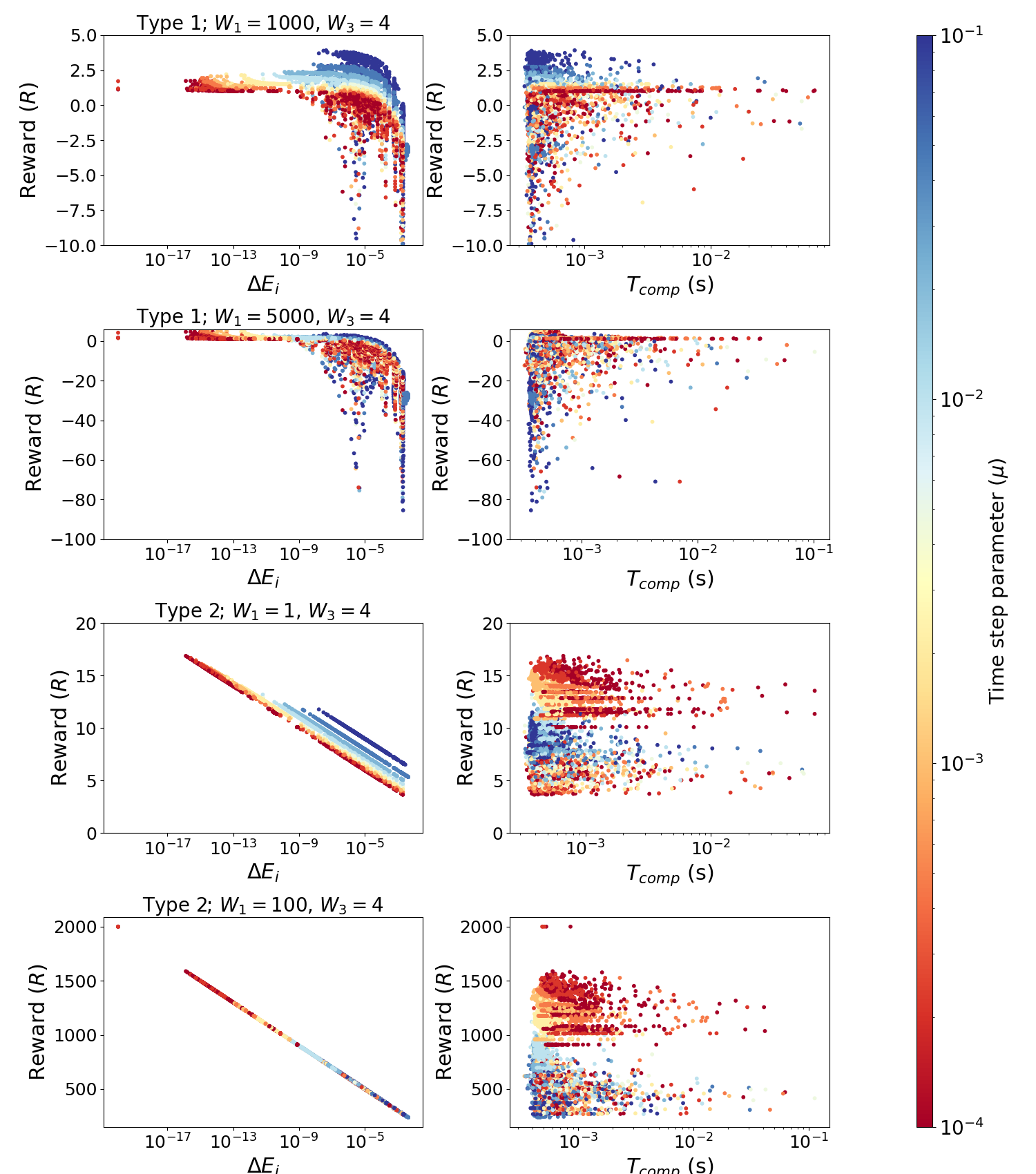}}
\caption{Reward value for different reward choices as in \autoref{eq:rewards} to \autoref{eq:reward4}. The left plot shows the reward as a function of $\Delta E_i$, and the right plot as a function  {of the computation time}. The color represents the size of the time step parameter.}
\label{fig:rewards_multiple}
\end{center}
\vskip -0.2in
\end{figure}

We show the behavior of each of those as a function of the energy error, and the  {computation time in} \autoref{fig:rewards_multiple}.  {Type 1 yields larger values of the reward for low energy errors and for high values of the time step parameter (shown in blue). The reward value rapidly decreases as the energy error rises to unphysical values. Type 2 presents a structure in which the maximum rewards are achieved by low energy errors, but a low computation time cannot lead to maximum values of the reward during the training procedure. }

 {For our training, we choose type 1} (\autoref{eq:rewards} and \autoref{eq:reward2}) since it shows the best adherence to our requirements.

\newpage
\section{Training and results for other initializations of Symple}
\label{appendix:resultsSymple}
\begin{figure}[h!]
	\vskip 0.2in
	\begin{center}
		\centerline{\includegraphics[width=0.6\columnwidth]{ 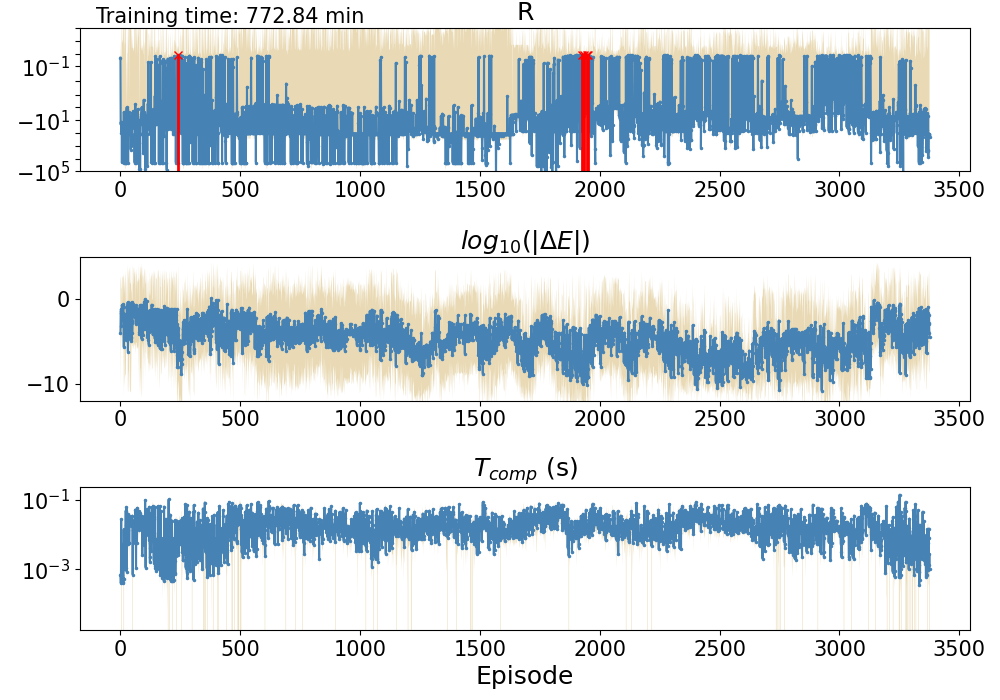}}
		\caption{Evolution of the average and standard deviation of different metrics of the test dataset per episode of: the reward value (first row), the energy error (second row), and the computation time (third row).}
		\label{fig:trainingSymple}
	\end{center}
	\vskip -0.2in
\end{figure}

We show the training evolution of the model for the {\tt Symple} integrator. We show the evolution of the reward, energy error, and computation time in \autoref{fig:trainingSymple}. The episodes with the largest reward values are indicated with red lines. We select the model at episode 1949. 

\begin{figure}[h]
	\begin{subfigure}[t]{0.49\textwidth}
		\includegraphics[width=\textwidth]{ 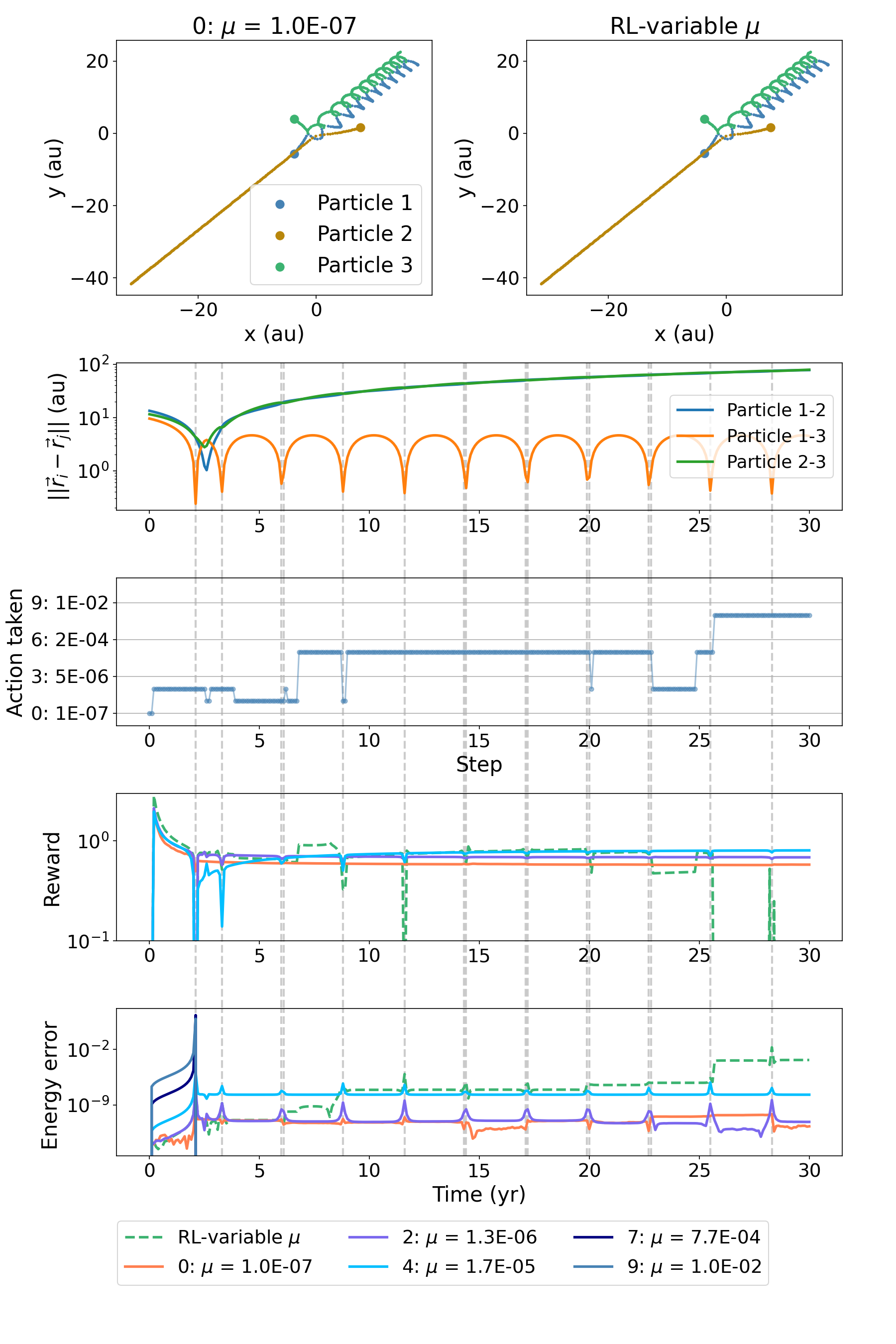}
		\caption{}
		\label{fig:comparisonRLvsfixedsizeSymple0}
	\end{subfigure}
	\begin{subfigure}[t]{0.49\textwidth}
	{\includegraphics[width=\textwidth]{ 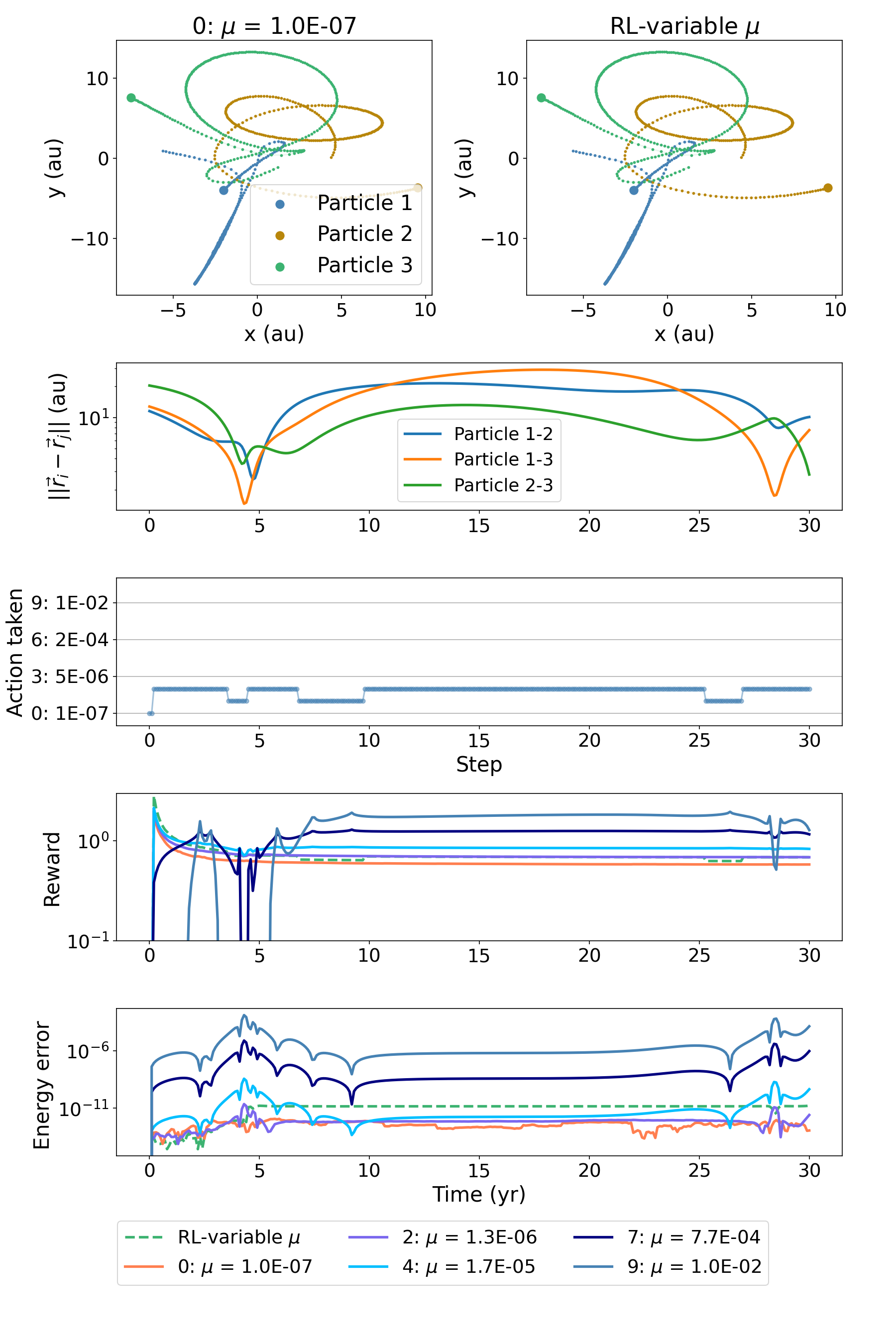}}
	\caption{}
	\label{fig:comparisonRLvsfixedsizeSymple2}
	\end{subfigure}
	\caption{Comparison of fixed-size time-step parameters with Reinforcement learning. We show the trajectory in Cartesian coordinates (top panel), the pairwise distance between particles (second panel), the action taken at each step (third panel), the reward for each case (fourth panel), and the energy error at each time step for each study case (fifth panel), for initializations with Seed 1 \textbf{(a)} and Seed 2 \textbf{(b)}.}
	
\end{figure}

We show other examples of the performance of our model trained for {\tt Symple}. In \autoref{fig:comparisonRLvsfixedsizeSymple0}, we see the evolution of the integration for the initialization with seed 1. As before, the model adapts the actions to keep the energy error constant  {while increasing the action number as the simulation progresses to improve the computation time.} In \autoref{fig:comparisonRLvsfixedsizeSymple2} we see similar results, with our model being able to keep the energy error constant.

\end{document}